\documentclass[pra,aps,amsmath,amssymb,superscriptaddress,longbibliography,notitlepage,twocolumn]{revtex4-1}
\usepackage{graphicx,multirow,outlines,ulem}
\usepackage{color}
\usepackage{hhline}
\usepackage{physics}
\usepackage{hyperref}
\usepackage{dsfont}

\usepackage[mathscr]{eucal}

\newcommand{\bla}{\color{black}}

\usepackage{tikz}
\usepackage{xcolor}

\newcommand{\bo}{\begin{outline}}
\newcommand{\eo}{\end{outline}}

\newcommand{\qed}{\nobreak \ifvmode \relax \else
      \ifdim\lastskip<1.5em \hskip-\lastskip
      \hskip1.5em plus0em minus0.5em \fi \nobreak
      \vrule height0.75em width0.5em depth0.25em\fi}

\begin{document}

\title{\textcolor{black}{Quest for
optimal quantum resetting: protocols for a particle on a chain}}
\author{Pallabi Chatterjee}
\email{ph22d001@iittp.ac.in } 
\author{S. Aravinda}
\email{aravinda@iittp.ac.in }   
\author{Ranjan Modak}
\email{ranjan@iittp.ac.in }
\affiliation{Department of Physics, Indian Institute of Technology Tirupati, Tirupati, India 517619}

\begin{abstract}
In the classical context, it is well known that, sometimes, if the search
does not find its target, it is better to start the process anew again, known as resetting. The quantum counterpart of resetting also indicates speeding up the detection process by eliminating the dark states, i.e., situations where the particle avoids detection. 
\textcolor{black}{
In this work, we introduce the most probable position resetting (MPR)
protocol in which, at a given resetting step, 
 resets are done with certain probabilities to the set of possible peak positions (where the probability of finding the particle is maximum) that could occur
because of the previous resets and followed by uninterrupted unitary evolution, irrespective of which path was taken by the particle in previous steps.} In a tight-binding lattice model, there exists a 2-fold degeneracy (left and right) of the
positions of maximum probability.
The survival probability with optimal restart rate approaches zero (detection probability approaches one) when the particle is reset with equal probability on both
sides path independently. This protocol significantly reduces the optimal mean first-detected-passage time (FDT) and performs better even if the detector is far apart compared to the usual resetting protocols where the particle is brought back to the initial position. We propose a modified protocol, an adaptive two-stage
MPR, by making the associated probabilities of going to the right and left a function of steps.  In this protocol, we see a further reduction of the optimal mean FDT and improvement in the search process when the detector is far apart. 
\end{abstract}
\maketitle

\section{Introduction}

In our daily life, we encounter many  stochastic resetting processes, such as animals
searching for food or persons searching for mobiles, that often include random resets in time, which involves say 
returning to the past locations where food was successfully located
or the last place a person remembers seeing her or his mobile. 
Usually, reset implies bringing back the system to its initial position, but in general, it can be at any other configuration  also~\cite{evans2011diffusion,evans2020stochastic}.  

{\bla
In the last few years, there has been a renewed interest in resetting protocols theoretically as well as experimentally,
given they can improve the efficiency of certain
random search processes and algorithms in the sense of reducing  mean
first-detection-passage time (FDT)~\cite{PhysRevLett.118.030603,pal.2019,PhysRevE.103.052129,evans2011diffusion,evans2020stochastic,gupta2014fluctuating,biswas2023rate,biswas2023stochasticity,pal2023random,ray2019peclet,tucci2022first,ali2022asymmetric,ray2022expediting,besga2020optimal,tal2020experimental,faisant2021optimal}.
In the context of the biochemical reaction, nature
found a way to accelerate the process with
a natural restart strategy~\cite{reaction1,reaction2,reaction3,reaction4}. \textcolor{black}{DNA-DNA hybridization, antigen-antibody binding, and various other molecular processes can all be described by the renowned Michaelis-Menten reaction scheme. Restart (or unbinding) is an integral part
of the renowned Michaelis-Menten reaction scheme~\cite{reaction1,reaction2,reaction3,reaction4}.} Reset processes have also
been studied in the context of non-equilibrium statistical mechanics, as they
evolve into a nontrivial non-equilibrium
stationary state~\cite{majumdar.2016,gupta.2017,majumdar.2015,majumdar2015dynamical,evans2013optimal,eule2016non,pal2015diffusion} also in the context of non-Markovian systems\cite{kumar2023universal}. There are many works regarding the effect of resetting on the ergodicity of several processes in the classical context\cite{wang2021time,vinod2022nonergodicity,wang2022restoring}.}

While stochastic resetting has been extensively studied in classical
non-equilibrium physics, its quantum counterpart is a relatively unexplored area\cite{PhysRevB.98.104309,PhysRevE.98.022129,sevilla2023dynamics,magoni2022emergent,10.21468/SciPostPhys.13.4.079,PhysRevB.104.L180302,Belan_2020}. Quantum restart has been studied using repeated projective measurements but without invoking any extra classical restart moves~\cite{kulkarni2023first}. On the other hand, the optimal rate of resetting has been calculated for a quantum walker on lattice subjected to repetitive projective measurements
on a particular lattice site in the presence of an additional classical move~\cite{yin2023instability,yin2023restart}, and a quantum advantage has been reported over classical resetting protocols~\cite{yin2023restart}.   

It has been shown that quantum restarts can help to overcome the dark subspace, which is caused by destructive interference~\cite{eli.2020,caruso2009highly,yin2023restart}. \textcolor{black}{An initial state that is never detected is called a dark state with respect to the detection state. For a finite-size system (ring), one can have this kind of dark eigenstates for which we do not get detection. On the other hand, for infinite
chains (the problem we are considering), \textcolor{black}{dark states remain dark. However, the bright states of the finite ring, while remaining orthogonal to every dark state, are not detected with unit probability in the case of the infinite chain. Instead, they are “dim”, such that $0 < P_{det} < 1$}}\cite{eli.2020}. Restart helps to speed up quantum search processes by removing the dark states.
Without resetting apriori, one would have expected that ballistic transport~\cite{PhysRevLett.100.013906} in non-interacting quantum systems should
act as an advantage in quantum search problems, but it turns out to be the opposite due to the presence of those dark sub-spaces. Hence, quantum resetting is fundamentally different than its classical counterpart. For example, the effects like the absence of continuous time measurement (quantum Zeno effect: the freezing of quantum dynamics)~\cite{misra1977zeno} and generation of entanglement due to quantum resetting~\cite{kulkarni2023generating} have no classical analog. In the quantum context, it has been studied that one can also use a very different approach, namely designing special graphs, and because of the constructive interference, one can find very fast first passage time\cite{liu2023designing}.

\begin{figure*}
    \centering
  \includegraphics[width=1.0\textwidth, height=0.8\textwidth]{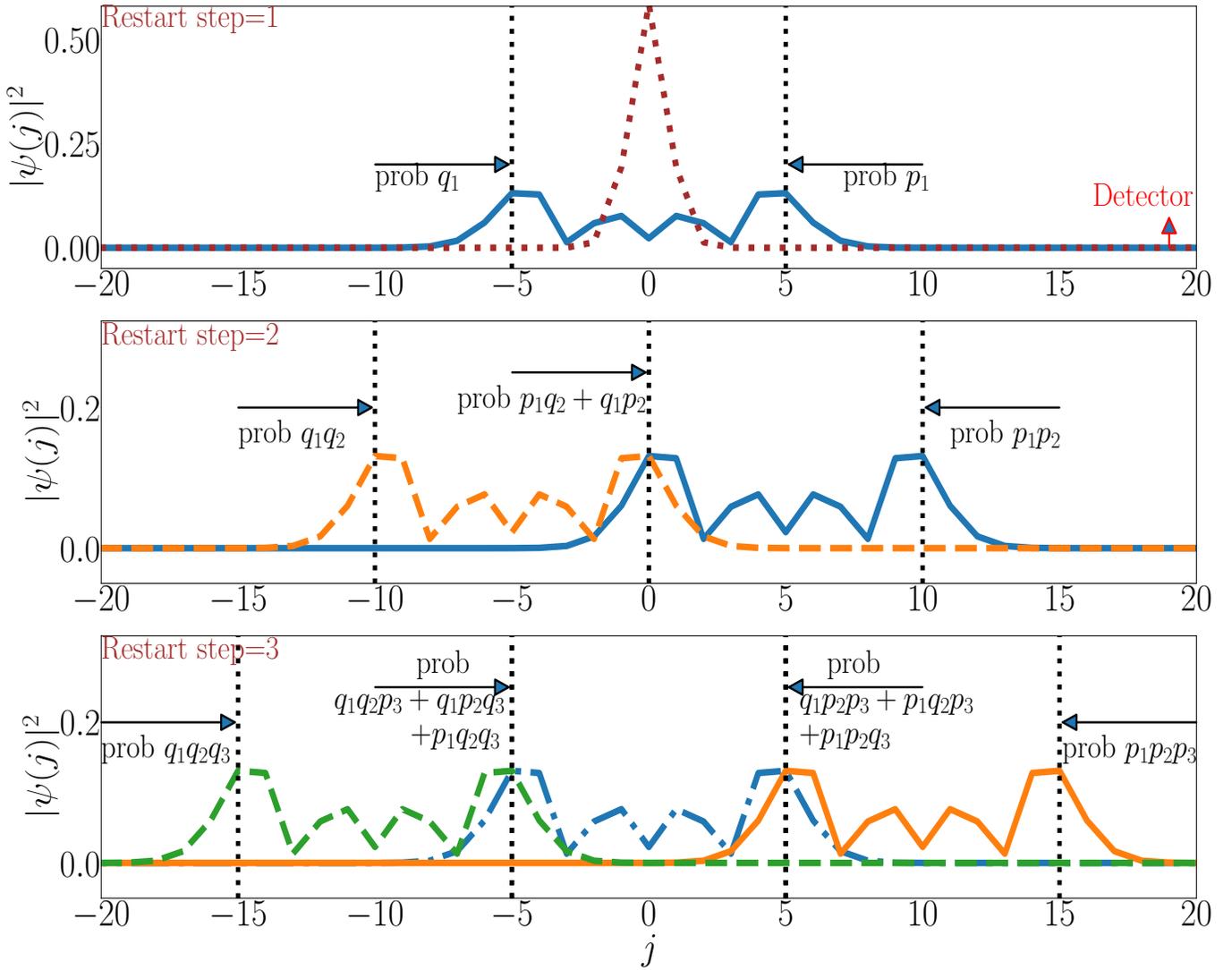}
    \caption{\textcolor{black}{Schematic of our proposed strategy based on the real data. First, we initialize the particle at $j=0$ site and let the wave function evolve for the time $t_r=3$ without any measurement interruptions. We see two peaks at $j=5$ and $j=-5$. Hence, we reset at positions $j=+5$ or $j=-5$. Now, as we have degeneracy in the reset positions, we associate probabilities $p_1$ and $q_1$ for reaching those right and left resetting positions by random walk starting from $j=0$. So, for the first resetting step, we reset either at $j=+5$ with probability $p_1$ or at $j=-5$ with probability $q_1$ (upper panel). Now, when we reset for the $2$nd time, we do not carry the history of the $1$st reset. Starting from $j=+5$, the next peaks are at $j=0,10$, and if it starts at $j=-5$, the next peaks are at $j=0,-10$. We have a total of three possible resetting positions at $j=0,+10,-10$ (middle panel). We associate the probabilities of choosing those peaks as going to those peak positions starting from $j=0$ by two steps doing a random walk, where the $1$st random walk step has the probability of going left and right as $q_1$ and $p_1$, respectively, and the $2$nd step has the probability $q_2$ and $p_2$ respectively. Hence, these three peaks at $j=\{-10,0,+10$\} are having the resetting probabilities $\{q^2, 2pq, p^2\}$. Similarly, in the $3rd$ step, we have $4$ peaks at $j=-15,-5,5,15$. We don't carry the memory of the previous steps and simply associate the probability of choosing the peaks as going to those peak positions from $j=0$ by random walk using three steps, where the third step of the random walk has the probability of going left and right as $q_3$ and $p_3$ (lower panel).  We consider $p_1=p_2=p_3=\cdots=p$ for the MPR protocol, while for the adaptive MPR protocol, we choose 
     $p_1=p_2=\cdots=p_{R_c}=p_I$ and $p_{R_c+1}=p_{R_c+2}=\cdots=p_F$.}}
    \label{fig0}
\end{figure*}

In this work, our goal is to find out the optimal resetting schemes under repeated monitoring of a particle moving on a tight-binding lattice that can be used as an efficient protocol for the search process. In this context, it is essential to point out that there has been a plethora of work on the quantum walk on tight binding lattice without resetting~\cite{tb1,tb2,tb3,tb4,tb5,tb6,modak2023non,thiel2018spectral}. {\bla We propose a scheme according to which we do not reset the particle to its initial position; instead, at a given resetting step, 
 resets are done with certain probabilities to the set of possible peak positions (where the probability of finding the particle is maximum) that could occur because of the previous resets and followed by uninterrupted unitary evolution, irrespective of which path was taken by the particle in previous steps, called the most probable position resetting (MPR).  As an example, we demonstrate our protocols in the form of a schematic based on the real data in Fig.~\ref{fig0}.}  We like to emphasize that in the case of the unbiased classical random walk, the probability distribution will always have a peak in the initial position; hence, the quantum counterpart of the protocol is the same as the initial position resetting (IPR)~\cite{yin2023restart}.  However, there is no classical counterpart to the MPR protocol. We find that the MPR protocol reduces the mean FDT significantly in comparison to the previously studied IPR protocols for optimal resetting time. Moreover, we introduce a two-stage adaptive most probable position resetting (Adaptive two-stage MPR) protocol by carefully varying the probability of going left and right in each step. Further reduction in mean FDT is observed in this protocol. The optimal mean FDT increases linearly with the increasing distance of the detector from the center. The slope of this linear dependence is smaller for MPR and adaptive MPR than the previously studied resetting protocols~\cite{yin2023restart}, which signifies the advantage of MPR and adaptive two-stage MPR over IPR when the detector is far apart from the center.

The manuscript is organized as follows. We introduce the tight-binding model and repetitive measurement formalism in Sec.\ref{sec:intro}. In Sec.\ref{sec:protocols}, we introduce different resetting schemes. Results are discussed in Sec.\ref{sec:results} and followed by discussions in Sec.\ref{sec:discussion}.

\section{Tight-binding model and repetitive measurement formalism \label{sec:intro}}

We consider a system which is governed by the following 
nearest-neighbour tight-binding Hamiltonian on a one-dimensional lattice of $2L+1$ sites, {\bla
\begin{equation}
    H=\sum_{j=-L}^{L-1}({c}^{\dag}_j{c}_{j+1}+\text{H.c.}),
    \label{Hamil}
\end{equation}}
where ${c}^{\dag}_j$ and ${c}_{j}$ are real-space fermionic creation and annihilation operators respectively. Under periodic boundary conditions, this Hamiltonian can be diagonalized easily by going into momentum space, and the Hamiltonian reads as 
$H=\sum_k \epsilon_{k}{c}^{\dag}_k{c}_{k}$, where $\epsilon_{k}=2\cos k$, ${c}^{\dag}_k$ and ${c}_{k}$ are momentum-space fermionic creation and annihilation operators. In this work, we focus on open boundary conditions, in particular, for all our results, we ensure that the wavefront does not reach the boundary so that all the results presented in our manuscript are valid for an infinitely long one-dimensional lattice.  

{\bla
Under the Hamiltonian evolution, the probability of finding the particle at a site 
$\delta$ (at state $\ket{\delta}$) after the time $t$ starting from an initial site $x_0$ (at state $\ket{x_0}$) is given as 
\begin{equation}
    \begin{split}
        P(\delta,x_0,t) & ={|\langle{\delta}|\psi(t)\rangle|}^2  \\
        &={|\langle{\delta}|\hat{U}(t)|x_0\rangle|}^2
\label{eqn 2}
    \end{split}
\end{equation}
where $U(t)=e^{-iHt}$ is the unitary operator. For a tight-binding  Hamiltonian $H$ in Eqn.~(\ref{Hamil}), it can be calculated as~\cite{PhysRevLett.120.040502}, the probability $P(\delta,x_0,t)$ is given as 
\begin{equation}
    P(\delta,x_0,t) = {|J_{|\delta-x_0|}(2t)|}^2,
\end{equation}
where $J_{n}(x)$ is the Bessel function of the first kind.

We would like to calculate first-detection probability at certain site under repetitive projective measurements~\cite{dhar2015quantum,PhysRevE.95.032141,tb3}. Note that due to the projective measurements, the evolution turns non-unitary. Consider a projective measurement at the site $\delta$ in the interval $\tau, 2\tau, \cdots, n\tau $, and the evolution is governed by the unitary $U(\tau)$ in between the measurements, and let the particle starts at $\ket{x_0}$. The state of the particle just before ($\ket{\psi_b}$)  and after(if the measurement does not detect the particle) ($ \ket{\psi_a}$) the  measurement at $t=\tau$, is 
\begin{equation}
    \begin{split}
        \ket{\psi_b} &= U(\tau) \ket{x_0} \\
        \ket{\psi_a} &= C (I-D) \ket{\psi_b} \\
        & = C (I-D)U(\tau) \ket{x_0}.
    \end{split}
\end{equation}
Here $C$ is the normalization constant and $D =\ketbra{\delta}$ is the projector onto the detection site. \textcolor{black}{In Appendix.~\ref{Appendix-C} we have plotted the square of the norm of the unnormalized wavefunction($|\psi_n^+\rangle$) just after $n=24(\tau=0.25)$ numbers of measurements (given measurements does not detect the particle)}. The probability of finding the particle at $\delta$ at time $\tau$ is $P_1 = \abs{\braket{\delta}{\psi_b}}^2$ and the probability of not detecting is $1-P_1$. Continuing the same way, we obtain the probability of detecting the particle at site $\delta$ at time $n\tau$ conditioned on no previous detection in the previous $n-1$ attempts   is 
\begin{equation}
    P_n = \frac{\abs{\langle\delta|\hat{U}(\tau)[(
I-\hat{D})\hat{U}(\tau)]^{n-1}|x_0\rangle}^2}{(1-P_1)(1-P_2) \cdots (1-P_{n-1})}.
\end{equation}
The probability of detecting  a particle at site $\delta$ for the first time at time $n\tau$ is thus given by 
\begin{equation}
    F_n = (1-P_1)(1-P_2) \cdots (1-P_{n-1}) P_n. 
\end{equation}
From this the probability amplitude for detecting the particle for the first time at time $n\tau$ in site $\delta$ can be defined as~\cite{PhysRevE.95.032141}
\begin{multline}
~~\phi_n^{x_0}=\langle\delta|\hat{U}(\tau)[(
\mathds{1}-\hat{D})\hat{U}(\tau)]^{n-1}|x_0\rangle
\\=\langle\delta|\hat{U}(n\tau)|x_0\rangle-\sum_{m=1}^{n-1}\langle\delta|\hat{U}[(n-m)\tau]|\delta\rangle\phi_m^{x_0}. 
\end{multline}

The probability of not detecting the particle at $\delta$ in the first $n$ attempts is called survival probability and is given by,
\begin{eqnarray}
    S_n=||[(
\mathds{1}-\hat{D})\hat{U}(\tau)]^{n}|x_0\rangle||^2
    =1-\sum_{m=1}^{n}F_m,
\end{eqnarray}
as, in terms of survival probability, the probability of first detection in the $n-th$ attempt, $F_n$, can be written as\cite{tb3},
\begin{equation}
    F_n=S_{n-1}-S_n.
\end{equation}
and, the total probability that the particle is detected at site $\delta$  in  $n$ attempts is 
\begin{equation}
    P_{det}(n) = \sum_{m=1}^{n}F_m = 1-S_n
\end{equation}
}

For the tight-binding Hamiltonian system, as $n\to \infty$, $P_{det}$ does not reach $1$, but it saturates to a value smaller than $1$ \cite{PhysRevLett.120.040502,yin2023restart}.
Interestingly, it turns out that even if $x_0=\delta$, $P_{det}(n\to\infty) < 1$ ~\cite{lahiri2019return}.  

It is important to point out that classically, the continuous time measurement 
process $\tau\to 0$ makes sense, but within the quantum framework, this leads to the freezing of the dynamics, which is known as 
 the quantum Zeno effect~\cite{misra1977zeno}. For all our calculations, we consider $\tau=0.25$, $|x_0\rangle=|j=0\rangle$, and $\delta=10$. We have examined the robustness of our main finding for different sets of parameters (see Appendix.~\ref{tau dependence}), though it is important to emphasize that in this work we have restricted ourselves to a small $\tau$ regime.    
 
\section{Resetting Protocols \label{sec:protocols}}
Repeated projective measurement without resetting cannot guarantee the detection in a quantum system, while resetting strategies can help ensure the detection. We discuss three different types of resetting protocols: {\bla Initial position resetting (IPR) $\mathcal{S}_{x_0}$ introduced in Ref~\cite{yin2023restart}, and two new protocols proposed by us called, most probable position resetting (MPR)  $\mathcal{S}^{p}_{x_M}$ and adaptive two-stage MPR $\mathcal{S}^{p_I,p_F}_{x_M}$ respectively.} 

\subsection{Initial Position Resetting (IPR) protocol $\mathcal{S}_{x_0}$} 

Assume we initialize a system by placing a particle at the site $x_0$. The dynamics are governed by the Hamiltonian $H$ and the repeated projective measurement at site $\delta$. We use the resetting scheme according to which one resets the system by repeatedly bringing the particle back to its initial position, i.e., at site $x_0$ after each time interval, resetting time $t_r=r\tau$, {\bla where  $r$ is a positive integer representing  the period of resetting called resetting rate}.  Depending on the value of resetting rate $r$,  it has been observed that $P_{det}(n\to\infty) \to 1$ and it guarantees the detection\cite{yin2023restart}. In the rest of the manuscript, we would refer to this protocol as the IPR strategy $\mathcal{S}_{x_0}$.

{\bla This protocol involves calculating  the first-hitting probability $F_n(r)$ after $n$ steps, in the presence of sharp resetting~\cite{PhysRevLett.118.030603}, where $r$ stands for resetting rate. Hence, one can write, $n=rR+\Tilde{n}$, where $R$=\text{integer}[($\frac{n-1}{r}$)] denotes the total number of resets before the detection. The first hitting probability in presence of resetting $F_n(r)$ can be obtained as ~\cite{yin2023restart},
\begin{equation}
    F_n(r) =[1-P^{x_0}_{det}(r)]^R F_{\Tilde{n}}^{x_0},
    \label{Eli's F_n}
\end{equation}
where ${P^{x_0}_{det}(r)}=\sum_{n=1}^r F_n^{x_0}$, \textcolor{black}{$F_n^{x_0}$ represents the first detection probability starting from the site $x_0$. The above Eq.~\eqref{Eli's F_n} represents the probability of no detection in a reset period($r$) to the power of the
number of reset periods($R$), combined with the probability of detection
after the remaining number of attempts $\Tilde{n}$ following the final reset.}

The total probability of detection up to time $n\tau$ in case of the IPR protocol $\mathcal{S}_{x_0}$ is,
\begin{equation}
    P^r_{det}(n)=\sum_{m=1}^n F_m(r),
    \label{pdet}
\end{equation}
and the corresponding survival probability is,
\begin{equation}
    S^r_n=1-P^r_{det}(n).
\end{equation}
The mean FDT under restart is 
\begin{equation}
{\langle t_f\rangle}_r=\langle n\tau\rangle=\tau\sum_{n=1}^{\infty}nF_{n}(r).
\label{t mean}
\end{equation}
Substituting the expression of $F_n(r)$ in Eq.~\eqref{t mean}, one  gets~\cite{yin2023restart,PhysRevLett.118.030603,PhysRevE.103.052129},
\begin{equation}
{\langle t_f\rangle}_r=\frac{r\tau[1-P^{x_0}_{det}(r)]}{P^{x_0}_{det}(r)}+\sum_{\Tilde{n}=1}^r\frac{\Tilde{n}\tau F^{x_0}_{\Tilde{n}}}{P^{x_0}_{det}(r)}.
\label{t mean x0}
\end{equation}}


\subsection{Most Probable Position Resetting(MPR) protocol $\mathcal{S}^{p}_{x_M}$}
{\bla In this section, we propose a resetting protocol where, instead of resetting to the initial position, i.e., at the site $x_0$ every time, in each resetting step, 
 resets are done with certain probabilities to the set of possible peak positions that could occur
because of the previous resets and followed by uninterrupted unitary evolution, irrespective of which path was taken by the particle in previous steps.}
   We start from an initial state $|x_0\rangle$. {\bla  If the wave function is allowed to evolve unitarily under the Hamiltonian $H$ for time $t_r=r\tau$,  the probability of finding the particle at any site $j$ at time $t_r$ is  $P(j,x_0,r\tau)$ (see Eq.~\eqref{eqn 2}).} In that case, the site where the particle most likely will be after time $t_r$ corresponds to the value of $j$ for which $P(j,x_0,r\tau)$ is maximum. 
We identify that site as $j^*_1$, and can write, 
\begin{equation}
    j_1^*=\max_j(P(j,x_0,r\tau)).\nonumber
\end{equation}
In the next steps, one has,
\begin{equation}
j_2^*=\max_j(P(j,j_1^*,r\tau)), \cdots ,j_R^*=\max_j(P(j,j_{R-1}^*,r\tau).\nonumber
\end{equation}


 For the {\bla tight-binding } Hamiltonian $H$ (where all the sites are identical), it turns out that if the resetting time $t_r$ is larger than some critical value (see Appendix.~\ref{Appendix-F}), the peak position of the traveling wave front will always be doubly degenerate, while below the critical $t_r$,  $j^*_k=x_0$ $\forall k\in [1, R]$.
For example, for a given $t_r$, if 
\begin{equation}
j_1^*=\{x_0+\Delta,x_0-\Delta\}, \nonumber
\end{equation}
then, 
\begin{equation}
\begin{split} 
    j_2^* &=\{\{x_0+2\Delta,x_0\},\{x_0,x_0-2\Delta\}\} \\
    j_3^*& =\{\{x_0+3\Delta,x_0+\Delta\},\{x_0+\Delta,x_0-\Delta\}\\
    &\{x_0+\Delta,x_0-\Delta\}, \{x_0-\Delta,x_0-3\Delta\}\}\\
&\vdots \\ 
j_R^*&=\{[x_0+(R-2m)\Delta]\binom{R}{m}\}, \text{where} ~~0\le m
\le R, m \in \mathbb{Z}. \\ 
    \label{eqn 6}
\end{split}
\end{equation} 
\textcolor{black}{The above equation represents, at the $R$'th resetting step we have possible resetting positions at [$x_0+(R-2m)\Delta$](occurring $\binom{R}{m}$ times),
where $m=0,1, \cdots R$, the set of all possible resetting positions at $R$-th resetting step, we are referring as $j_R^*$, and here $\Delta=|j^*_1-x_0| \geq 0$. Note that for $\Delta=0$, which corresponds to very small
$t_r$, the protocol is the same as the IPR protocol $S_{x_0}$.} \textcolor{black}{At a given resetting step, we reset at any of the possible resetting positions from the set $j_R^*$ irrespective of the fact that where it was reset in the previous steps, that makes our protocol path-independent. Also, we associate probabilities for choosing a reset position at a given resetting step $R$ from the set of available resetting position $j_R^*$ as the probability of going to that particular position doing random walk by $R$ step starting from $x_0$. For the MPR protocol, for every random walk step, we associate probabilities $p$ and $q$ for right and left, respectively.} \textcolor{black}{It means if we initialize the particle at $x_0$, then in the 1st step we reset the particle at  $x_0+\Delta$ with probability $p$
and at  $x_0-\Delta$ with probability $q$, further in the 2nd step 
we reset the particle at  $x_0+2\Delta$ with probability $p^2$,
at  $x_0-2\Delta$ with probability $q^2$, and at $x_0$ with probability 
$2pq$, and so on (see Fig.~\ref{fig0}). Important to emphasize that in our protocol, 
we are not keeping track of the resetting history of the particle in the previous steps, e.g., in the 2nd step of our protocol, in order to reset the particle at $x_0+2\Delta$ ($x_0-2\Delta$); it is not necessary that previously the particle was reset at site $x_0+\Delta$ ($x_0-\Delta$).  Independent of where the particle was reset in the 1st step, in the 2nd step, we reset them at $x_0+2\Delta$ with probability $p^2$,
at  $x_0-2\Delta$ with probability $q^2$, and at $x_0$ with probability 
$2pq$.
We have also studied a scenario where we track the previous resetting trajectories of the particle (which we refer to as path-dependent MPR protocol); results remain qualitatively the same as what we find here, which are described in  Appendix.~\ref{Appendix-E}.
\textcolor{black}{However, note that the path-dependent MPR protocol is difficult to analyze theoretically as compared to the path-independent one. Hence, we have restricted ourselves to the path-independent protocols in the main text.}}

\textcolor{black}{If in the whole process, we restart the system $R$ times in the positions taken from the sets $j_1^*$, $j_2^*$,......$j_R^*$ respectively, where $j_R^*$ represents the set of all possible resetting positions at $R$-th resetting step, then one can modify Eq.~\eqref{Eli's F_n} according to MPR protocol $S^p_{x_M}$ and can write,
\begin{equation}
    F_n(r)=\prod_{i=0}^{R-1}[1-\Tilde{P}^{i}_{det}(r)]\Tilde{F}_{\Tilde{n}}^{R},
    \label{new F_n}
\end{equation}
where ${\Tilde{P}^{i}_{det}(r)}=\sum_{n=1}^r \Tilde{F}_n^{i}$, and
\begin{equation}
\Tilde{F}_{\Tilde{n}}^{R}=\sum_{l=0}^R  |\phi_{\Tilde{n}}^{x_0+(R-2l)\Delta}|^2\binom{R}{l}p^{R-l}(1-p)^l.
\label{binom}
\end{equation}}
\textcolor{black}{$\Tilde{F}_{{n}}^{R}$ represents the first detection probability in the $n-th$ measurement, summing over the contributions of starting from all possible sites in the $R-th$ resetting step with the corresponding probability of being that particular site.}

{\bla Let us validate the above expressions Eq.~\eqref{new F_n} and Eq.~\eqref{binom} with an example of $n=3,r=1,R=2$.
For the $1st$ resetting step, possible peaks are at $x_0+\Delta$ and $x_0-\Delta$, and associated resetting probabilities are $p$ and $q$, respectively. In the $2nd$ step, the possible peaks are at $x_0+2\Delta, x_0, x_0-2\Delta$ with resetting probabilities $p^2, 2pq, q^2$ respectively independent of where the particle was reset in the 1st step. Hence, in the expression of $F_3(1)$, there will be contributions from $6$ possible paths: 1) resetted at $x_0+\Delta$ in the 1st step, and then in the 2nd step at 
$x_0+2\Delta$, 2) resetted at $x_0+\Delta$ in the 1st step, and then in the 2nd step at 
$x_0$, 3) resetted at $x_0+\Delta$ in the 1st step, and then in the 2nd step at 
$x_0-2\Delta$. 4) resetted at $x_0-\Delta$ in the 1st step, and then in the 2nd step at 
$x_0+2\Delta$, 5) resetted at $x_0-\Delta$ in the 1st step, and then in the 2nd step at 
$x_0$, and 6) resetted at $x_0-\Delta$ in the 1st step, and then in the 2nd step at 
$x_0-2\Delta$. Mathematically, it reads as, }
{\bla
\begin{align*}
F_3(1)=&(1-|\phi_1^{x_0}|^2)p(1-|\phi_1^{x_0+\Delta}|^2)p^2(|\phi_1^{x_0+2\Delta}|^2)\\&+(1-|\phi_1^{x_0}|^2)p(1-|\phi_1^{x_0+\Delta}|^2)2pq(|\phi_1^{x_0}|^2)\\&+(1-|\phi_1^{x_0}|^2)p(1-|\phi_1^{x_0+\Delta}|^2)q^2(|\phi_1^{x_0-2\Delta}|^2)\\&+(1-|\phi_1^{x_0}|^2)q(1-|\phi_1^{x_0-\Delta}|^2)p^2(|\phi_1^{x_0+2\Delta}|^2)\\&+(1-|\phi_1^{x_0}|^2)q(1-|\phi_1^{x_0-\Delta}|^2)2pq(|\phi_1^{x_0}|^2)
\\&+(1-|\phi_1^{x_0}|^2)q(1-|\phi_1^{x_0-\Delta}|^2)q^2(|\phi_1^{x_0-2\Delta}|^2)\\
=&(1-|\phi_1^{x_0}|^2)(1-p|\phi_1^{x_0+\Delta}|^2-q|\phi_1^{x_0-\Delta}|^2)\\&~~~\cdot\big{[}p^2(|\phi_1^{x_0+2\Delta}|^2)+2pq(|\phi_1^{x_0}|^2)+q^2(|\phi_1^{x_0-2\Delta}|^2)\big{]}\\
=&(1-\Tilde{P}^{0}_{det}(1))(1-\Tilde{P}^{1}_{det}(1))\Tilde{F}_1^2
\end{align*}
}
{\bla This is identical to what one obtained from the Eq.~\eqref{new F_n}. See Appendix.\ref{Appendix-D} for the numerical validation in case of large $R$. 
On the other hand, for the path-dependent protocol, one would have contributions from only 4 paths:  1) resetted at $x_0+\Delta$ in the 1st step, and then in the 2nd step at 
$x_0+2\Delta$, 2) resetted at $x_0+\Delta$ in the 1st step, and then in the 2nd step at 
$x_0$, 3) resetted at $x_0-\Delta$ in the 1st step, and then in the 2nd step at 
$x_0$, and 4) resetted at $x_0-\Delta$ in the 1st step, and then in the 2nd step at 
$x_0-2\Delta$. Hence, mathematically, $F_3(1)$ for path-dependent protocol will read as,
\begin{align*}
F_3(1)=&(1-|\phi_1^{x_0}|^2)p(1-|\phi_1^{x_0+\Delta}|^2)p(|\phi_1^{x_0+2\Delta}|^2)\\&+(1-|\phi_1^{x_0}|^2)p(1-|\phi_1^{x_0+\Delta}|^2)q(|\phi_1^{x_0}|^2)\\&+(1-|\phi_1^{x_0}|^2)q(1-|\phi_1^{x_0-\Delta}|^2)p(|\phi_1^{x_0}|^2)
\\&+(1-|\phi_1^{x_0}|^2)q(1-|\phi_1^{x_0-\Delta}|^2)q(|\phi_1^{x_0-2\Delta}|^2).
\end{align*}
}
{\bla Note that this expression is different from the results obtained for the path-independent protocol using Eq.~\eqref{new F_n}(see Appendix.~\ref{Appendix-E} for more details). Unlike path-independent protocol, it is not possible to get a nice expression like Eq.\eqref{new F_n} in the case of path-dependent protocol. That makes the path-independent protocol easier to analyze theoretically as compared to the path-dependent one.}\\
Inserting the modified expression of $F_n(r)$ into Eq.\eqref{pdet} and Eq.~\eqref{t mean}, one can obtain the 
total probability of detection $P^r_{det}$ up to time $n\tau$ and the mean FDT ${\langle t_f \rangle}_r$ for the protocol 
$\mathcal{S}^p_{x_M}$. \textcolor{black}{The results of the above expressions can be interpreted as, at a resetting step $R$, we
reset at any of the positions from set $j_R^*$ with the appropriate probability. Again in the next step, we reset at any of the possible positions from the set $j_{R+1}^*$ with the corresponding probability without tracking the previous reset trajectories.
We verify the correctness of the above equations by performing a numerical experiment (see Appendix~\ref{Appendix-D} for details).} 

It is important to point out that unlike in the case of $\mathcal{S}_{x_0}$, it is not possible to obtain an expression like Eq.~\eqref{t mean x0} for $\mathcal{S}^p_{x_M}$, we evaluate, ${\langle t_f\rangle}_r=\langle n\tau\rangle=\tau\sum_{n=1}^{n_c}nF_{n}(r)$. 
In general, ${\langle t_f \rangle}_r$ increases with $n_c$.  We fit the data with a polynomial of $1/n_c$. Given we are interested in the $n_c\to \infty$ result, one can read the intersection of the fitted polynomial with the $y$ axes as the extrapolated value of ${\langle t_f \rangle}_r$. (see Appendix.~\ref{extrapolation}).


\textcolor{black}{Finally, we summarize our protocols in $4$ simple steps, \\
\\
Step 1: We initialize the particle at the center of the lattice at $x_0$ and keep the detector at site $\delta$, which clicks in the interval $\tau, 2\tau, 3\tau, \cdots, n\tau,\cdots$.\\
\\
Step 2: To compute the first detection probability $F_n(r)$ in the $n$-th measurement, first we fix a resetting time $r\tau$, which automatically fixes the maximum number of resetting steps are allowed, i.e,  $R=int[\frac{n-1}{r}]$. Identify the distance $\Delta$ of the peaks in the time-evolved probability density from the initial position of the particle at time $r\tau$.  \\
\\
Step 3: 
Keep resetting the particles at time intervals of $r\tau,  
 2r\tau, \cdots, Rr\tau$, at  the following sites  $\{x_0+\Delta,x_0-\Delta\}, \{x_0+2\Delta, x_0, x_0-2\Delta\},\cdots,
   \{[x_0+(R-2m)\Delta]\}, \cdots$, where $0\le m
\le R, m \in \mathbb{Z}$ respectively with probabilities $\{p,q\}, \{p^2, 2pq, q^2\}, \cdots, \{\binom{R}{m}p^{R-m}(1-p)^m\}$. \\
\\
Step 4: Calculate the first detection probability $F_n(r)$ in the $n$-th measurement, taking the average over all possible realizations, mathematically, which is represented by Eq.\eqref{new F_n}.}

\subsection{Adaptive two-stage Most Probable Position Resetting (Adaptive two-stage MPR) protocol $\mathcal{S}^{p_I,p_F}_{x_M}$ }
 We introduce another protocol, which is a modified version of MPR protocol $\mathcal{S}^{p}_{x_M}$, where we had the probability $p$ ($q$)  of going at right (left) remains the same in each step. {\bla On the other hand, here we propose a scheme where the probability $p$ can also vary in each random walk step in a given resetting step, e.g. one can think of a situation where in the $1st$ resetting step, we have the probability of resetting at $x_0+\Delta$ and $x_0-\Delta$ as $p_1$ and $q_1$ respectively. In the $2nd$ resetting step, the possible positions are $x_0+2\Delta, x_0, x_0-2\Delta$. In this process, starting from $x_0$ (center of the lattice), we consider the probability of going to the left and right in the first random walk step is $p_1$ and $q_1$, and in the second random walk step, these probabilities become $p_2$ and $q_2$. The probability of reaching at $x_0+2\Delta, x_0, x_0-2\Delta$ are $p_1p_2, p_1q_2+q_1p_2, q_1q_2$ respectively. A similar rule is applicable for all the resetting steps (see fig. \ref{fig0}).} In this case, one can write,
\begin{equation}
\begin{split}
\Tilde{F}_n^{0} &=|\phi_n^{x_0}|^2 \\
\Tilde{F}_n^{1} &=p_1|\phi_n^{x_0+\Delta}|^2 +q_1|\phi_n^{x_0-\Delta}|^2 \\
\Tilde{F}_n^{2}&=p_1p_2|\phi_n^{x_0+2\Delta}|^2 +p_1q_2|\phi_n^{x_0}|^2 + \\
& q_1p_2|\phi_n^{x_0}|^2 + q_1q_2|\phi_n^{x_0-2\Delta}|^2 \\
& \vdots \\
\Tilde{F}_n^{R}& =\sum_{l=0}^R\frac{1}{l!}\bigg[\sum_{i_1,i_2,\cdots i_l=1,i_1\neq i_2\neq \cdots \neq i_l}^R q_{i_0}q_{i_1}q_{i_2}\cdots q_{i_l}\\
& \prod_{k=1,k\neq i_1 \neq i_2 \neq \cdots \neq \cdots i_l}^R p_k\bigg]|\phi_n^{x_0+(R-2l)\Delta}|^2 
\label{assymetric}
\end{split}
\end{equation} 
with $q_{i_0}=1$.
 
Given the numbers of individual terms increase exponentially with $R$ in Eq.~\eqref{assymetric}, here we consider a simpler scheme, i.e. we associate only two real numbers $p_I$ ($q_I=1-p_I$) and $p_F$ ($q_F=1-p_F$) with our scheme. {\bla Previously, for the MPR protocol, the probability of resetting to a possible peak at the $R-th$ resetting step was taken as the probability of reaching that peak in $R$ steps starting from the site $x_0$ doing a random walk. According to the adaptive two-stage MPR protocol, we consider that among those $R$ random walk steps, $R_c$ number of steps have probability $p_I$($q_I$) of going right(left), and $R-R_c$ steps have probability $p_F$($q_F$) of going right(left).}  One is free to choose the optimal value for $p_I$, $p_F$, and $R_c$ to speed up the detection process.  \textcolor{black}{We choose $p_F=0.5$ (the reason is discussed in the next section), and $p_I$ can be thought of as an adaptive /confidence parameter. Suppose, based on some previous knowledge, if one thinks that more likely the search object is on the right side of the lattice, $p_I$ can be chosen $>0.5$. Based on the confidence level, one will choose $p_I$ accordingly. }

Like before, here  we have a very similar expression of $F_n(r)$ as in Eq.~\eqref{new F_n}, but the expression of $\Tilde{F}^{R}_{\Tilde{n}}$ will be modified as,
\begin{equation}
\begin{split}
\Tilde{F}_{\Tilde{n}}^{R}= &\sum_{l=0}^{R_c} \binom{R_c}{l}p_I^{R_c-l}(1-p_I)^l\bigg[\sum_{m=0}^{R-R_c} \binom{R-R_c}{m}\\
& p_F^{R-R_c-m}(1-p_F)^m |\phi_{\Tilde{n}}^{x_0+(R-2l-2m)\Delta}|^2\bigg],\\ 
& \text{where} \quad R\ge R_c.
\label{R>Rc}
\end{split}
\end{equation} 
and,
\begin{equation}
\begin{split}
\Tilde{F}_{\Tilde{n}}^{R}=& \sum_{l=0}^R  |\phi_{\Tilde{n}}^{x_0+(R-2l)\Delta}|^2\binom{R}{l}p_I^{R-l}(1-p_I)^l,\\
& \text{when} \quad  R< R_c.
\label{R<Rc}
\end{split} 
\end{equation}
Total probability of detection $P^r_{det}$ up to time $n\tau$ and the mean first hitting time ${\langle t_f \rangle}_r$ can be calculated from  Eq.~\eqref{pdet} and Eq.~\eqref{t mean} respectively using the modified expression of $F_n(r)$.

\begin{figure}
    \centering
    \includegraphics[width=0.465\textwidth, height=0.3\textwidth]{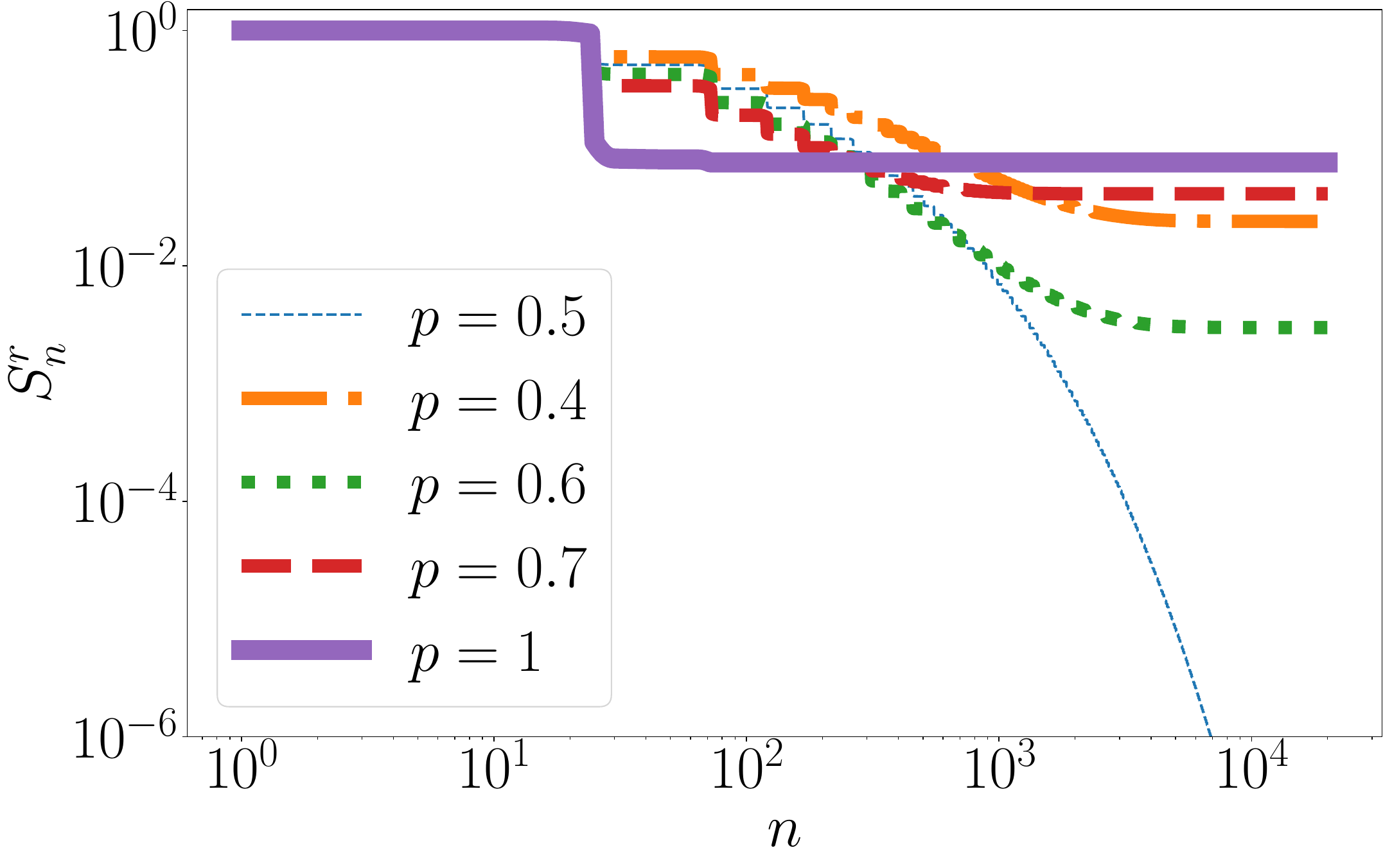}
    \caption{Dependence of Survival probability $S_n^r$ on $n$ in presence of restart, for different values of $p$. Results are for the protocol $\mathcal{S}^p_{x_M}$, and  and $r=24$.}
   \label{fig:1}
\end{figure}

\begin{figure}
    \centering
    \includegraphics[width=0.465\textwidth, height=0.3\textwidth]{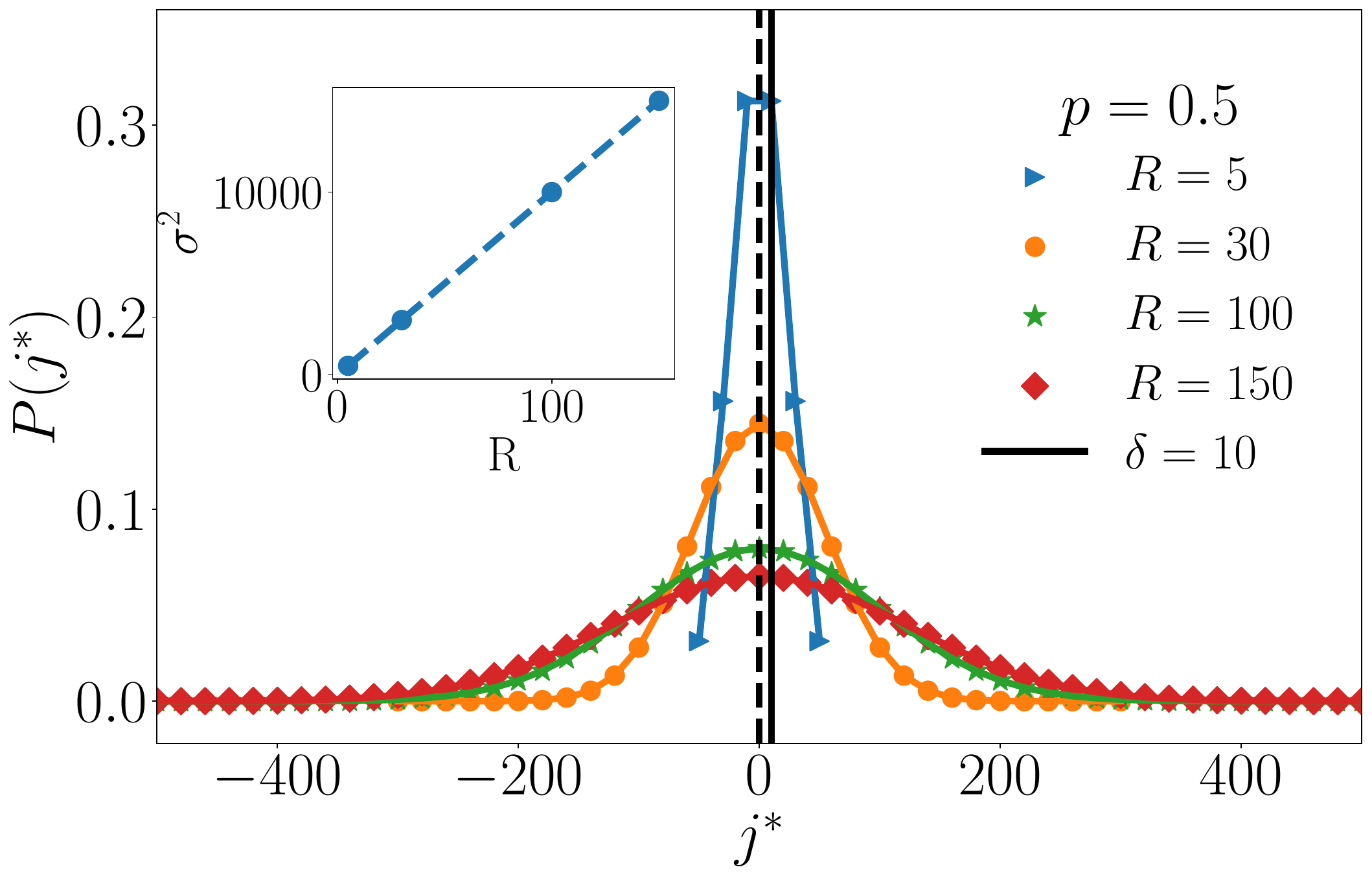}
    \label{fig:enter-label}
    \includegraphics[width=0.465\textwidth, height=0.3\textwidth]{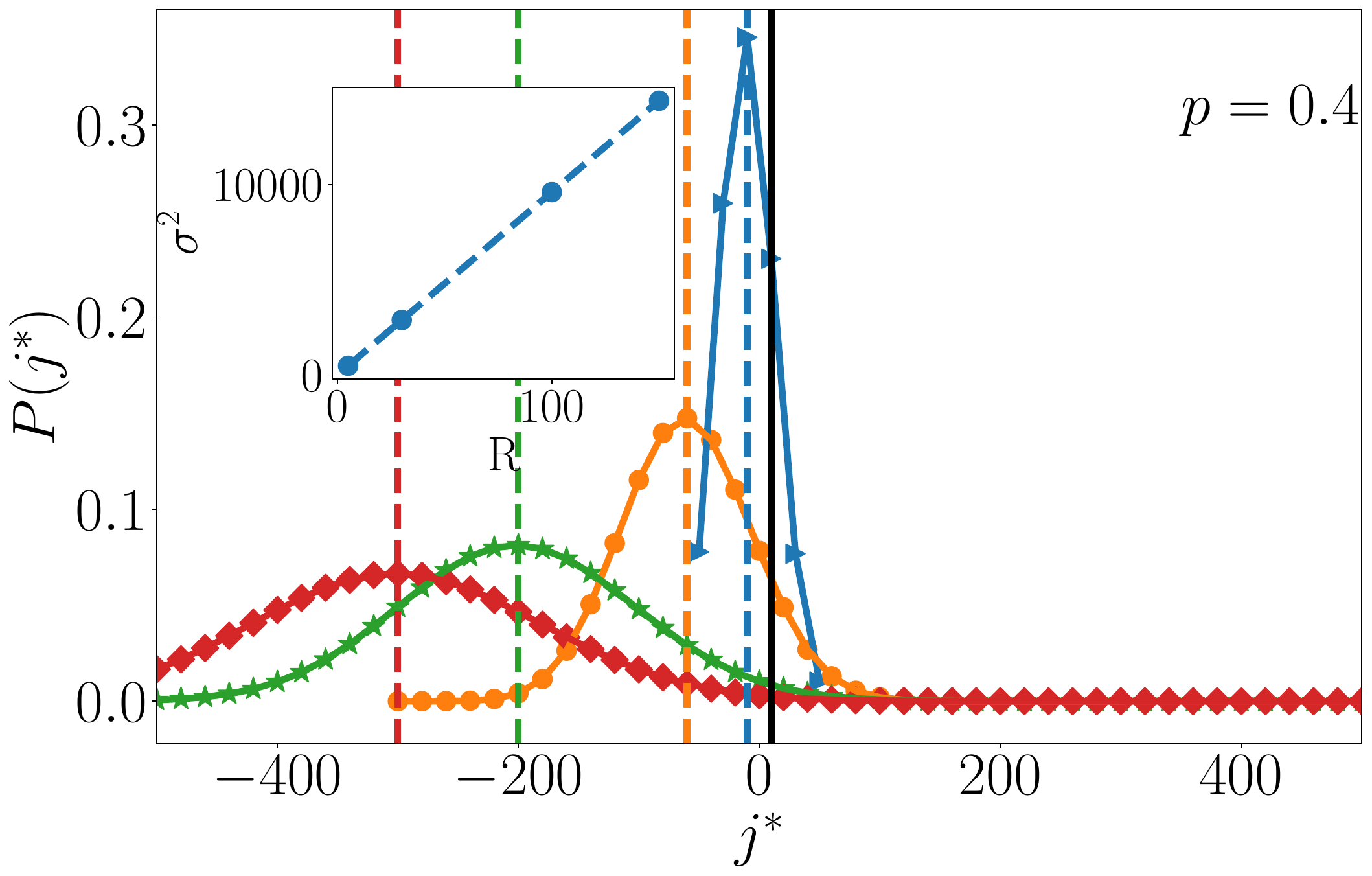}
    \includegraphics[width=0.465\textwidth, height=0.3\textwidth]{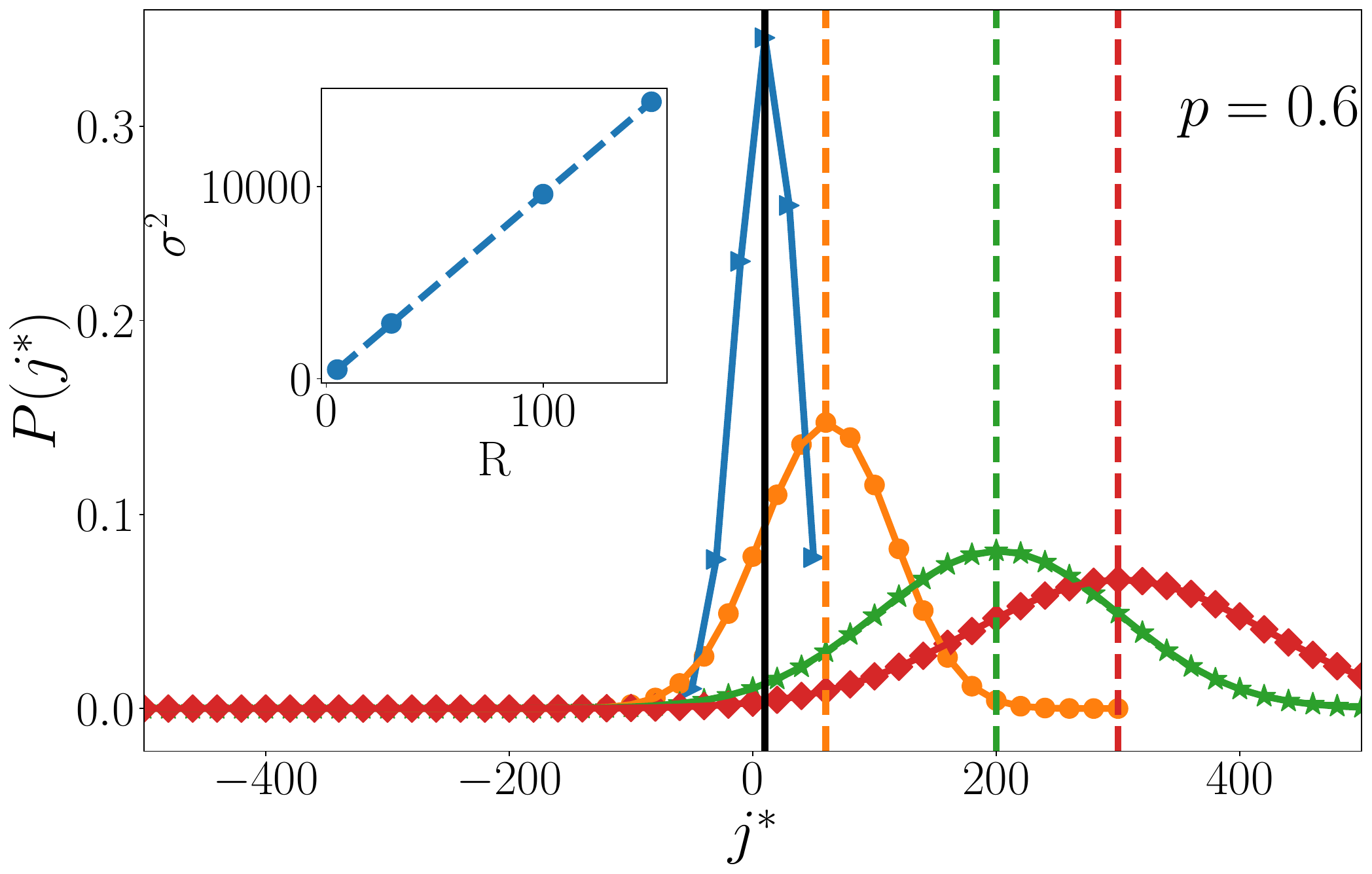}
    \caption{Probability distribution of $j^*$  for different values of $p$ (upper panel: $p=0.5$, middle panel: $p=0.4$, lower panel: $p=0.6$) and different number of resetting steps $R$ under the ${\mathcal{S}}^p_{x_M}$ protocol. The dashed line shows the position of $\langle j^*_R \rangle$, and the solid line shows the position of the detector. Inset shows $R$ dependency of the variance ${\sigma}^2_R$ of the distribution. Results are for  $r=24$.}
    \label{fig:2}
\end{figure}

\section{Results \label{sec:results}}
Given we know the large $n$ behavior of the survival probability or total detection probability for the IPR protocol, we compare these quantities with our newly proposed  MPR and two-stage Adaptive MPR protocols. We will see that for the MPR protocol, survival probability saturates to a value greater than $0$ for $p=0.4,0.6,0.7,1$ and approaches $0$ for $p=0.5$. We connect the reason for the saturation of the survival probability with the probability distribution of the $j^*$. We see a great reduction in mean FDT for the MPR protocol over the IPR protocol for optimal restart time. Later, we see that for the adaptive two-stage MPR protocol, the mean FDT for the optimal restart rate is the lowest among all three protocols. Interestingly, we see even if the optimal mean FDT increases linearly with the increasing distance of the detector from the center; the slope is much smaller for our proposed resetting protocols MPR and two-stage Adaptive MPR than the previously studied resetting scheme IPR, which implies protocols MPR and two-stage Adaptive MPR are even better than IPR when the detector is far apart.  Below we explain all our results in detail.

\subsection{Probability distribution of $j^*$ and its connection to the saturation of survival probability for MPR protocol} 

We want to check the fate of survival probability $S_n^r$ for the protocol MPR. 
Figure.~\ref{fig:1} clearly demonstrates that for $p=0.4,0.6,0.7,1$, survival probability $S^r_n$ saturates to a non-zero value, which automatically implies  $P^r_{det}(n\to\infty) <1$. This implies the detection is not guaranteed for these $p$ values.
On the other hand, in the case of $p=0.5$, one can see that $S^r_n$ is approaching zero in the large $n$ limit. Hence, one can expect a complete detection $P^r_{det}(n\to\infty) \to 1$.

We will now try to explain the above-mentioned nature of $S^r_n$ for different $p$ values. The probability distribution of $j^*$ (i.e., the set of possible resetting positions at a given resetting step) is completely analogous to the probability distribution of the position of a one-dimensional biased discrete random walk (see Eq.~\eqref{eqn 6}). The interval $\Delta$ and the number of resetting steps $R$ in our strategies are analogous to the step length and the number of total jumps associated with a random walk problem. In the random-walk picture, the $p$ ($1-p$) value can be thought of as the probability of going to the right (left) sites.   
Drawing the mapping between a random walk problem and MPR protocol can help us to compute the mean and variance of the $P(j^*)$, which are given by,  
\begin{equation} 
\begin{split}
\langle j^*_R\rangle &= R\Delta(2p-1),\\
     \sigma^2_R &=4R\Delta^2p(1-p)
     \label{variance}
\end{split}
\end{equation}

Figure.~\ref{fig:2} shows the probability distribution of $j^*$ for different $p$ values and for different number of resetting steps $R$. For $p=0.5$ (Fig.~\ref{fig:2} upper panel) the mean of the distribution $P(j^*)$ always remains at $j=0$. The black dashed line, and the black solid line in Fig.~\ref{fig:2} (upper panel) represent the mean of the distribution $P(j^*)$ and the detector position, respectively.
On the other hand, for $p>0.5$ (see lower panel of Fig.~\ref{fig:2}) and $p<0.5$ (see middle panel of Fig.~\ref{fig:2}) the mean of the distribution $P(j^*)$ is heading towards the right and left side respectively with increasing $R$, as one can expects from the Eq.~\eqref{variance} as well. The shifting of the mean $\langle j^*_R \rangle$ with increasing $R$ can be seen by tracking the dashed vertical lines that represent the mean of the distribution. While in Fig.~\ref{fig:2} middle panel for $p=0.4$ those dashed lines move toward left, in the lower panel for $p=0.6$ they move towards right with increasing $R$. 
In all these cases, the variance $\sigma_R^2$  also increases with $R$ linearly, hence obeying the prediction of Eq.~\eqref{variance} (see insets of Fig.\ref{fig:2}). 

It is important to point out that $\langle j^*_R\rangle\propto R$ and $\sigma_R\propto R^{1/2}$. In the large $R$ limit, $P(j^*)$ becomes a Gaussian distribution with $\langle j^*_R\rangle >> \sigma_R$. Naively, one would expect that given $\delta>0$, the detection will be faster if $p> 0.5$. 
However, as the probability distribution of $j^*$ in the neighborhood of $\delta$ (which also indirectly contributes to the $P_{det}$ computation) will be almost negligible when $\langle j^*_R\rangle -2 \sigma_R >> \delta$, $P_{det}$ is not expected to grow with $R$ anymore beyond a certain point, and it saturates to some finite value smaller than $1$. 


\begin{figure}
    \centering
    \includegraphics[width=0.465\textwidth, height=0.3\textwidth]{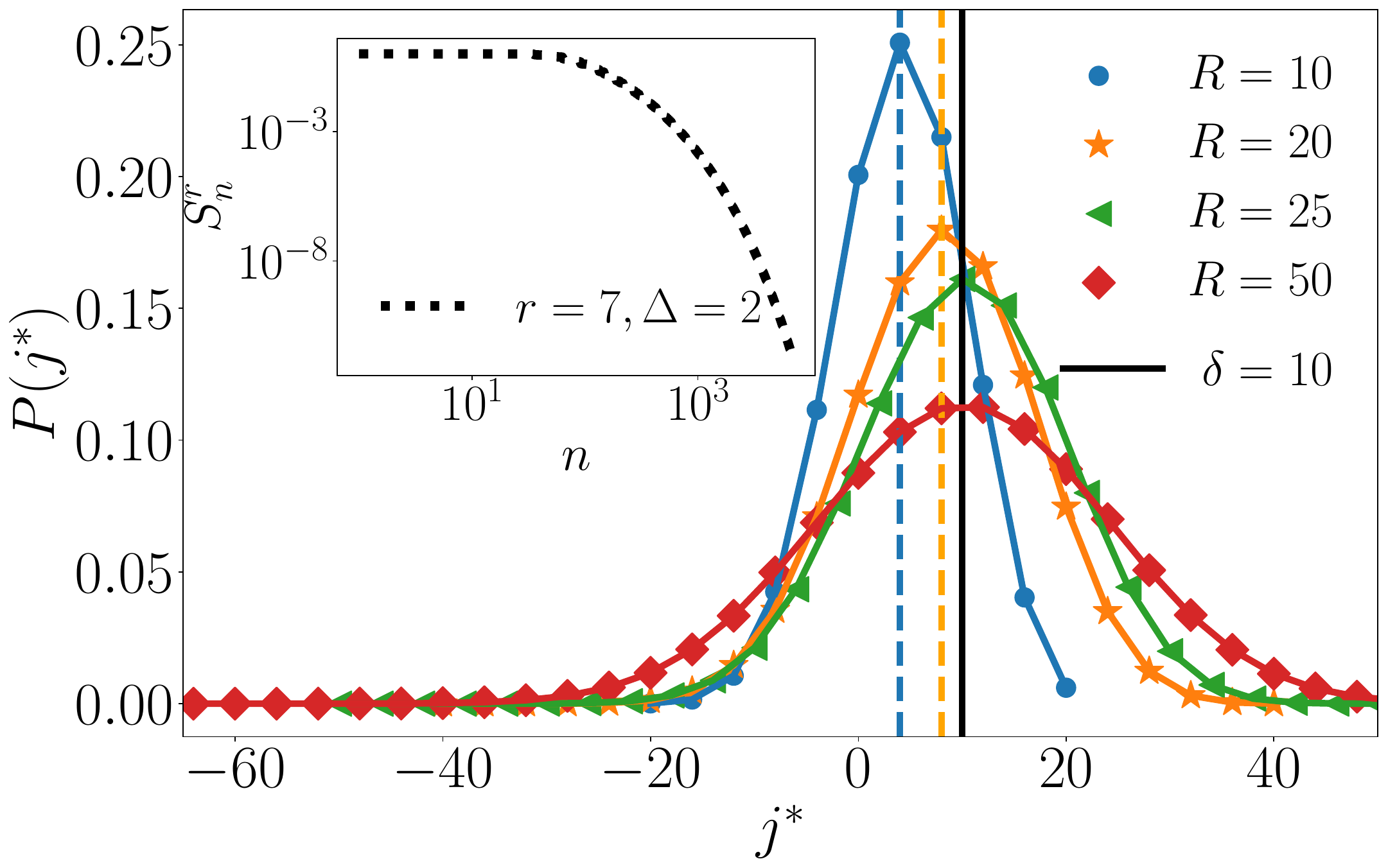}
    \caption{Probability distribution of $j^*$ under $\mathcal{S}^{p_I,p_F}_{x_M}$ protocol for different values of $R$. Dashed lines represent the position of $\langle j^*_R \rangle$ and the solid line represents the position of the detector. Results are for $r=7$, $p_I=0.6$, $p_F=0.5$. Inset shows the $n$ dependence of survival probability $S^r_n$.}
    \label{fig:assym}
\end{figure}

\subsection{Survival probability and probability distribution of $j^*$ for Adaptive two-stage MPR protocol}

In this section, we present the results for the  adaptive two-stage MPR protocol with $p_I=0.6$ and $p_F=0.5$. \textcolor{black}{Given $p_F$ is chosen to be $0.5$, the peak position or the mean of $j^*$ distribution will not change with $R$ for $R>R_c$; however, the location of the peak will depend on the choice of $R_c$. One would expect that if we could choose $R_c$ such that the peak/mean of the $P(j*)$ remains near the detector site $\delta$, that will enhance the possibility of detection more.}
Hence, in our calculations, we use,
\begin{equation}
    R_c=[\frac{\delta}{\Delta(2p_I-1)}],
    \label{Rc eqn}
\end{equation}
where, $[]$ denotes the nearest integer, e.g. $[5.873]=6$
and $[5.173]=5$.
\textcolor{black}{Note that in the subsequent section, we have also considered the scenario where we have chosen $p_I$ and $R_c$ randomly.}
Inset of Fig.~\ref{fig:assym} clearly shows that the survival probability($S^r_n$) is approaching zero with increasing $n$. So, in this case, as $P^r_{det}(n)$ approaches one with increasing $n$, one can also expect complete detection. Figure.~\ref{fig:assym} (main panel) shows the probability distribution $P(j^*)$.
The results are for $r=7$, and the $R_c$ is chosen according to Eq.~\eqref{Rc eqn}. Figure.~\ref{fig:assym} shows that for $R \leq R_c$ mean $\langle j^*_R \rangle$ of the distribution $P(j^*)$ is shifting towards right obeying Eq.~\eqref{variance},
for $R>R_c$ mean does not shift anymore. The dashed lines in Fig.~\ref{fig:assym} represent the shifting of the $\langle j^*_R \rangle$ towards the right. Because of this particular choice of $R_c$, $\langle j^*_R \rangle$ gets fixed very close to the detector (represented by the black solid line in Fig.~\ref{fig:assym}), and with increasing $R$, the variance increases but the mean does not move any further. 

For the  MPR protocol with $p=1/2$ and two-stage adaptive MPR, the main commonality is that in both cases, the mean of the distribution $P(j^*)$ does not move indefinitely with increasing $R$. However, while in the former protocol, the mean is fixed at its initial position, in the case of the latter, it gets fixed at a site in the neighborhood of the detector site $\delta$ (sometimes even at the $\delta$ site as well). 
Hence, we predict that the  adaptive two-stage MPR protocol is much more efficient in improving the efficiency of the search process. 


\subsection{Comparison of mean FDT for IPR, MPR, and Adaptive two-stage MPR protocols}

\begin{figure}
    \centering
    \includegraphics[width=0.465\textwidth, height=0.3\textwidth]{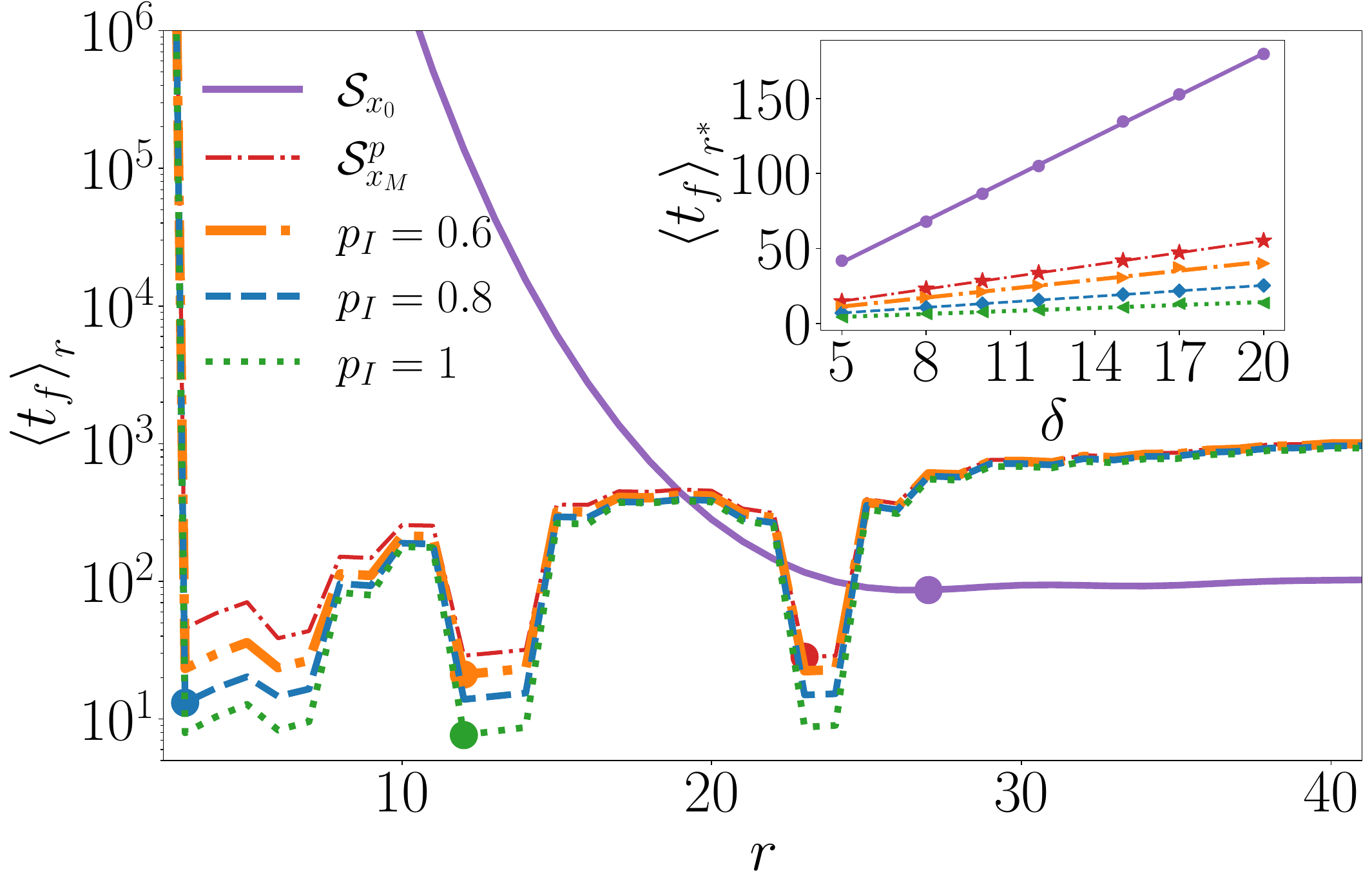}
    \caption{Main Panel: Mean first hitting time ${\langle t_f\rangle}_r$ vs. $r$ plots for three protocols($\mathcal{S}_{x_0}$, ${\mathcal{S}}^p_{x_M}$, $\mathcal{S}^{p_I,p_F}_{x_M}$) for $\delta=10$. Circles represent the min(${\langle t_f\rangle}_r$).
    Inset: Mean first hitting time for the optimal restart rate $r^*$(${\langle t_f\rangle}_{r^*}$) vs. $\delta$ plots for three protocols $\mathcal{S}_{x_0}$, ${\mathcal{S}}^p_{x_M}$, and, $\mathcal{S}^{p_I,p_F}_{x_M}$.}
    \label{t-mean}
\end{figure}

We compute the mean first detection time (FDT) for these three protocols and compare them. Figure.~\ref{t-mean} shows the value of mean FDT for different restart rates $r$. Minima of the ${\langle t_f \rangle}_r$ curve with respect to $r$ gives the optimal restart rate $r^*$, we identify them by solid circles in the plot. From Fig.~\ref{t-mean} it is clear that,
\begin{equation}
{\langle t^{\mathcal{S}^{p_I,p_F}_{x_M}}_f \rangle}_{r^*} < {\langle t^{{\mathcal{S}}^p_{x_M}} \rangle}_{r^*}< {\langle t^{\mathcal{S}_{x_0}} \rangle}_{r^*}.
\end{equation}
It is confirmed that our proposed protocols, MPR and two-stage adaptive MPR, both are doing a better job of reducing the optimal mean FDT, hence much more efficient compared to the protocol IPR. For protocols MPR and two-stage adaptive MPR, one sees multiple local minimas that correspond to  $r$ for which $\frac{\delta}{\Delta} \in \mathbb{Z}$. For example, given our choice of $\delta=10$, we observe $4$ local minima and they correspond $\Delta=1$, 2, 5, and 10 respectively. 
It turns out that among the protocols IPR, MPR, and adaptive 
 two-stage MPR protocol, adaptive two-stage MPR 
with $p_I=1$ and $p_F=0.5$ is the most efficient one, given  
$\langle t_f \rangle_{r^{*}}$ is the minimum. Note while it is clear from our results (also kind of expected) that minimum mean FDT will correspond to one of such $\Delta$s so that $\frac{\delta}{\Delta} \in \mathbb{Z}$, but it is not obvious that which one will correspond to the actual minimum mean FDT, we find different results for different values of $p_I$ for the protocol adaptive two-stage MPR. 

Inset of Fig.~\ref{t-mean} shows the dependence of optimal mean FDT on $\delta$, and though it is linear in the case of all the three protocols IPR, MPR, and Adaptive two-stage MPR, the slope is much less for MPR and Adaptive two-stage MPR protocols than IPR protocol. And as expected the slope is the least for the Adaptive two-stage MPR protocol. So the protocols MPR and Adaptive two-stage MPR are superior to IPR even if the detector is far apart.  We also like to point out that 
for protocols MPR and adaptive two-stage MPR, one can not obtain a simple expression for mean FDT for the infinite sum in Eq.~\eqref{t mean} (which one can for ${\mathcal{S}}_{x_0}$, see Eq.~\eqref{t mean x0}), hence, the
data presented in Fig.~\ref{t-mean} for protocols MPR and two-stage adaptive MPR are extrapolated results and the extrapolation technique has been discussed in Appendix.~\ref{extrapolation}.

\subsection{Comparison of total detection and survival probability up to time $n\tau$ for IPR, MPR and two-stage Adaptive MPR protocols}

\begin{figure}
    \centering
    \includegraphics[width=0.465\textwidth, height=0.3\textwidth]{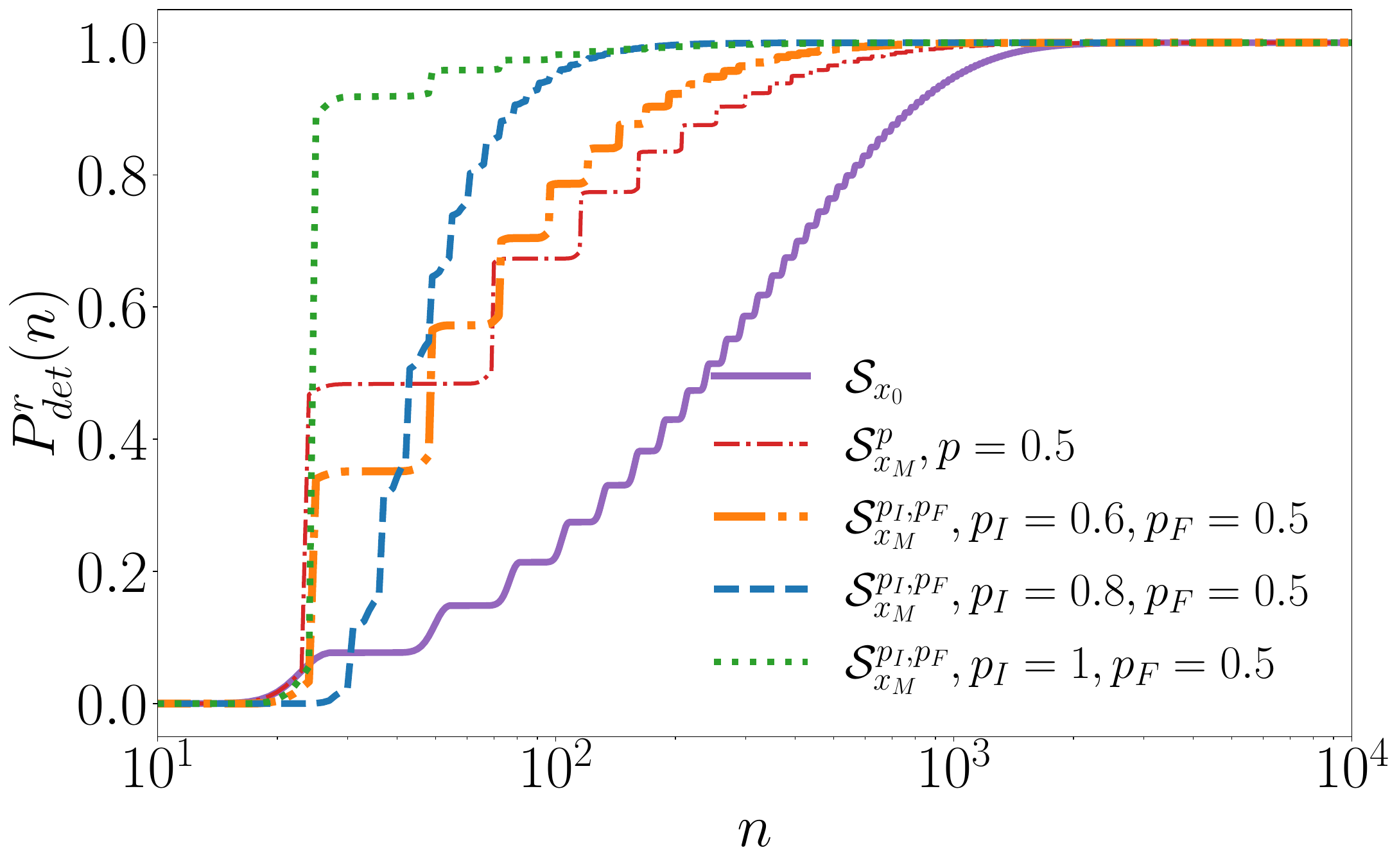}
    \caption{$P^r_{det}(n)$ vs. $n$ plots for the optimal resetting rate for all the three protocols $\mathcal{S}_{x_0}$, ${\mathcal{S}}^p_{x_M}$, and $\mathcal{S}^{p_I,p_F}_{x_M}$.  }
    \label{fig:pdet}
\end{figure}

 We want to compare $P^r_{det}$ for the optimal restart rate $r^*$ for all three protocols. Figure.~\ref{fig:pdet} represents how $P^r_{det}(n)$ grows with increasing $n$ and approaches one, for the optimal $r^*$. Figure.~\ref{fig:pdet} shows the initial growth of $P^r_{det}$ for protocols MPR and adaptive two-stage MPR are much faster than the protocol IPR (represented by the solid line). We find the fastest initial growth of $P^r_{det}$ for the protocol adaptive two-stage MPR with $p_I=1$ and $p_F=0.5$, which is represented by the green dotted line in Fig.~\ref{fig:pdet}.

\begin{figure}
    \centering
    \includegraphics[width=0.465\textwidth, height=0.3\textwidth]{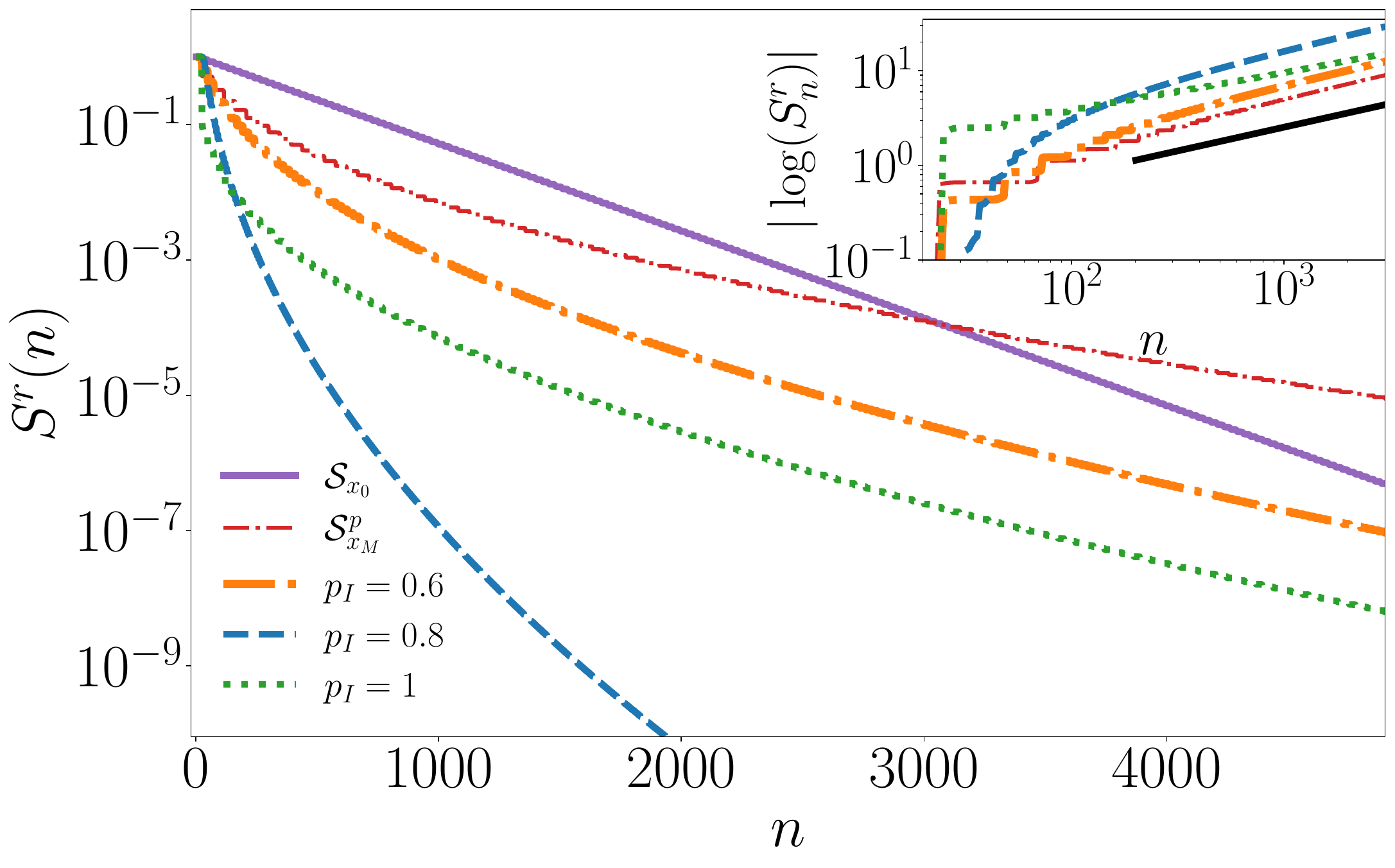}
    \caption{$S^r(n)$ vs. $n$ plots for the optimal resetting rate for all the three protocols $\mathcal{S}_{x_0}$, ${\mathcal{S}}^p_{x_M}$, $\mathcal{S}^{p_I,p_F}_{x_M}$. Inset shows $n$ dependency of $|\log(S^r_n)|$, which is linear in log-log scale for large $n$ limit. The black solid line is the reference line of slope $1/2$.  }
    \label{fig:survival}
\end{figure}

 We calculate the survival probability $S^r_n$ for all the protocols for optimal restart rate $r^*$. Figure.~\ref{fig:survival} shows in the case of MPR and adaptive two-stage MPR protocols,   $S^r_n$ decays much faster than IPR protocol (represented using the solid line in the main panel) up to a certain $n$. The initial decay rate is the maximum for the protocol two-stage adaptive MPR, with $p_I=1$, represented by the green dotted line. For the protocol IPR, the decay of $S^r_n$ is exponential, as one can see in Fig.~\ref{fig:survival} ($S^r_n$ is a straight-line in semi-log scale). We investigate the nature of the decay of $S^r_n$ for the other two protocols. Inset of Fig.~\ref{fig:survival} shows the dependency of $|\log(S^r_n)|$ on $n$ in a $\log-\log$ scale, and one can see in the large $n$ limit, $|\log(S^r_n)|$ is linear in $n$ with slope close to $1/2$ (the black solid line is a reference line for slope $1/2$). This indicates that the decay of $S^r_n$ is stretched exponential, i.e. $S^r_n\approx e^{-an^{1/2}}$ in contrast to the exponential decay has been observed for $\mathcal{S}_{x_0}$. Indeed in the large $n$ limit, the fall-off of $S^r_n$ with $n$ will be much faster for $\mathcal{S}_{x_0}$ compared to the other two schemes, which is also apparent from the main panel of Fig.~\ref{fig:survival}. In terms of mean FDT for optimal restart rate, the advantages of our proposed protocols are enormous in comparison to the initial position resetting scheme. 

\subsection{Calculations of mean FDT for finite number of measurements}\label{finite measurements}
\textcolor{black}{In the previous sections, we discussed the results of the two-stage adaptive MPR protocol, where we took $R_c$ such that the mean of the distribution of $j^*$ remains very close to the detector site $\delta$. 
We find that the mean FDT corresponds to the optimal resetting rate, which is reduced significantly depending on the choice of the adaptive parameter $p_I$. However, it involves optimizing three free parameters $(p_I, R_c, t_r)$. Hence, in this section, we show the results of the conditional mean FDT\cite{wang2024firstnew} for the finite number of measurements for the random choice of $R_c$ and even for random values of $p_I$. Here, we investigate two scenarios for a random choice of $R_c$; in one case, we choose $p_I$ randomly from the window $[0,1]$, and in another case, we choose $p_I>0.5$. As it is very difficult to get a convergent result averaging over $R_c$, $p_I$ with increasing $n$, we restrict ourselves to finite $n_c$ calculations. The conditional mean FDT is given by, ${\langle t_f \rangle}^{cond}(r,n_c)=\frac{\sum_{n=1}^{n_c}nF_n(r)}{\sum_{n=1}^{n_c}F_n(r)}$. It is important to emphasize the finite number of measurements is also experimentally realizable~\cite{wang2024firstnew}. In this finite number of measurements $n_c$, the maximum number of restarts that can happen is $R_{max}=integer[\frac{n_c-1}{r}]$. Hence, in each realization, we randomly choose  $R_c$ and finally average over many realizations.}

\begin{figure}[!h]
    \centering
    \includegraphics[width=0.465\textwidth, height=0.3\textwidth]{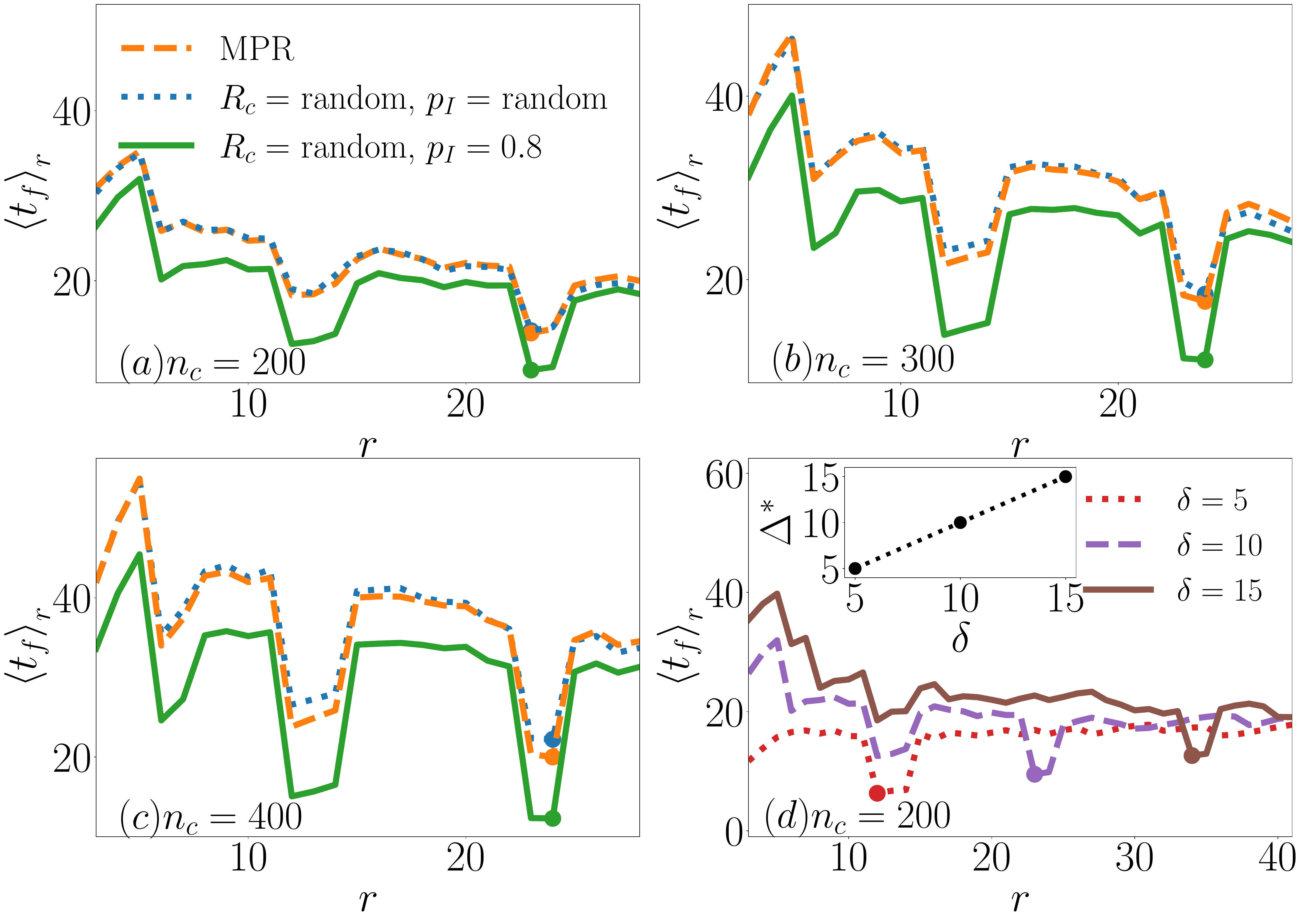}
    \caption{${\langle t_f \rangle}_r$ vs. $r$ plots for finite number of measurements $n_c=200$ ($a$), $n_c=300$ ($b$), $n_c=400$ ($c$). In (a)-(c) the dotted line represents the result with random $R_c$ and random $p_I$, the solid line represents the result with random $R_c$ and $p_I=0.8$, and the dashed line represents the result for MPR. (d) represents the ${\langle t_f \rangle}_r$ vs. $r$ plots for different $\delta$ values with $n_c=200$. Results are for random $R_c$ and $p_I=0.8$. Inset shows the optimal $\Delta^*$ scales linearly with $\delta$. Circles represent the optimal restart rates for the corresponding protocols.  }
    \label{finite plots}
\end{figure}

 \textcolor{black}{Fig.~\ref{finite plots} represents ${\langle t_f \rangle}_r$ vs. $r$ plots for finite number of measurements $n_c=200$ ($a$), $n_c=300$ ($b$), $n_c=400$ ($c$). In (a)-(c), the dotted line represents the result with random $R_c$ and random $p_I$, the solid line represents the result with random $R_c$ and $p_I=0.8$, and the dashed line represents the result for MPR. It shows that random $R_c$ and random $p_I$
 results are very similar to the MPR protocol. However, as soon as we introduce the adaptive parameter $p_I >0.5$ (given we are more than $50\%$ confident that the search object is at the right-hand side of the lattice), we find significant improvement if indeed the search object is at the right-hand side of the lattice.  
 (d) represents the ${\langle t_f \rangle}_r$ vs. $r$ plots for different $\delta$ values with $n_c=200$. Results are for random $R_c$ and $p_I=0.8$. Inset shows the optimal $\Delta^*$ scales linearly with $\delta$. Circles represent the optimal restart rates for the corresponding protocols. In all the cases, one can see the optimal $r^*$ is for which $\Delta=\delta$.  }

\section{Discussion \label{sec:discussion}}

For a classical particle, doing an unbiased random walk on a one-dimensional lattice where at a given site a detector is placed,  it is well established that the average hitting time or mean FDT reduces if one resets the particle on its original position again and again with an optimal time interval. Interestingly, if one computes the probability distribution of the sites visited by that classical stochastic particle, the mean and peak of the distribution both lie in the same position and that coincides with the particle's initial position as well. However, this is not the case for a quantum particle moving on a tight-binding lattice. As a matter of fact, it turns out that the mean of the distribution coincides with the initial position of the quantum particle, but the peak does not. On the other hand, the sites where the distribution is peaked signify where the particle is most likely to be found. {\bla According to MPR protocol, at a given resetting step, 
 resets are done with certain probabilities to the set of possible peak positions that could occur
because of the previous resets and followed by uninterrupted unitary evolution, irrespective of which path was taken by the particle in previous steps.} We have found that the efficiency of the search process increased significantly compared to the case where resetting is done at the initial position, with guaranteed detection when the associated probabilities of going to the right and the left at each random walk step required to reach a possible resetting position($R$ number of random walk steps at the $R$-th resetting step) are equal. We believe that our MPR resetting protocol can be a natural choice in the quantum context that has no classical analog. 
\textcolor{black}{We would like to point out that we had restricted ourselves in this work to the nearest-neighbor hopping model on a 1D chain, which is exactly solvable; hence, finding the most likely position is
straightforward. However, if we do not have a solvable system, one can numerically study the unitary evolution
to find out the most probable positions as a function of $t_r$ for such models. Hence, in principle, our study can
be extended even for non-solvable models with extra classical simulations. {\bla In that case, more rigorous and system-dependent studies are required to assure quantum advantage.}}

We also propose another protocol, which we call Adaptive two-stage MPR. One can think of a situation where the particle has a priori memory about the location of the target e.g., say the walker has a rough estimation about the distance between its initial position and the target and also the walker is relatively confident (confident level has been incorporated in the parameter $p_I$ in adaptive two-stage MPR) that whether the target is on the right side or the left side of its initial position. In those cases, one can use our adaptive two-stage resetting protocol with great success. We see the rate of the linear increment of the optimal FDT with $\delta$ is lesser in the case of MPR and Adaptive two-stage MPR protocols, which implies our proposed protocols are even more efficient than IPR if the detector is far apart from the center.

We believe our resetting protocols can be of great use in quantum search algorithms. \textcolor{black}{This kind of tight-binding Hamiltonians as well as many-body Hamiltonians(XXZ model) and their dynamics can be implemented in quantum computers using qubit representations and Trotterization\cite{smith2019simulating,ibm,IBMR,wang2024firstnew}. Moreover, such repeated measurements and the effect of resetting  have
been implemented on IBM quantum computers where they implemented the projective nonunitary dynamics
using a product of elementary gate operators with the help of Trotterization.~\cite{ibm,IBMR,wang2024firstnew} and our work will enable the way to speed up quantum hitting times on quantum computers.} Also, quantum walks are widely applied in many different fields, starting from transport in wave guides to ultra-cold atoms to light harvesting dynamics in biochemistry~\cite{PhysRevA.58.915,PhysRevA.48.1687,PhysRevLett.100.013906}. Hence, our proposed resetting schemes might have implications in all these different fields as well. While we have limited ourselves to the tight-binding lattice where all states are delocalized in this work, it will be interesting to understand the role of quantum resetting schemes for systems where localized and delocalized states can coexist~\cite{chatterjee2023one,deng2017many} and also for many-body interacting systems~\cite{modak2021finite}. It will be interesting to study the effect of our resetting protocols on the recently studied thermodynamics phases in the first detected return times of quantum many-body systems\cite{walter2023thermodynamic}.

\section{Acknowledgements}
RM acknowledges the DST-Inspire fellowship by the
Department of Science and Technology, Government of
India, SERB start-up grant (SRG/2021/002152). SA acknowledges the start-up research grant from SERB, Department of Science and Technology, Govt. of India
(SRG/2022/000467). The authors thank D Mondal and S Ray for fruitful discussions. 

\appendix

\begin{figure}
    \centering
    \includegraphics[width=0.465\textwidth, height=0.3\textwidth]{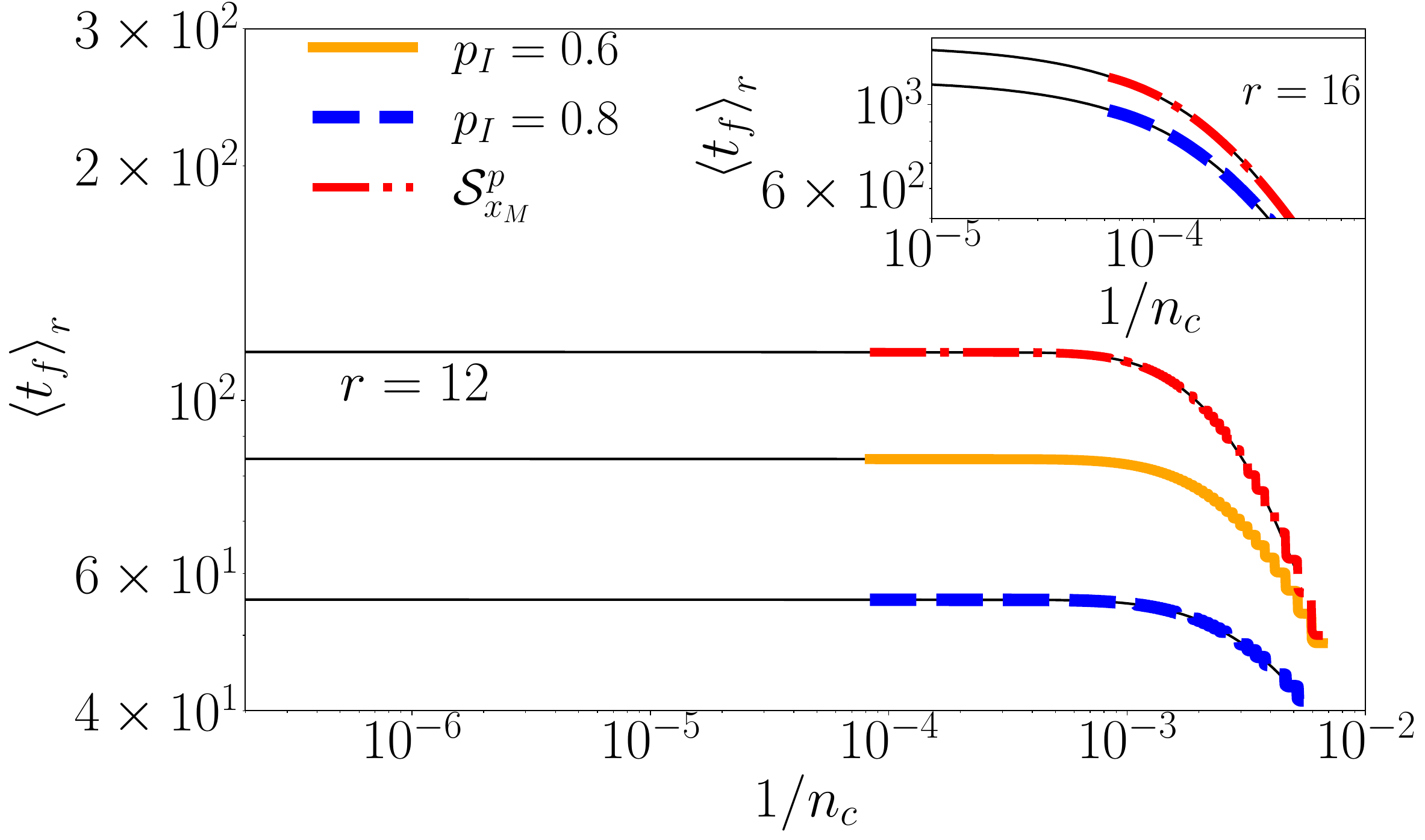}
    \caption{Main panel: ${\langle t_f \rangle}_r$ vs. $1/n_c$ plots for $r=12$ for protocols ${\mathcal{S}}^p_{x_M}$ and $\mathcal{S}^{p_I,p_F}_{x_M}$.
    Inset shows ${\langle t_f \rangle}_r$ vs. $1/n_c$ plots for $r=16$ for the protocols ${\mathcal{S}}^p_{x_M}$ and $\mathcal{S}^{p_I,p_F}_{x_M}$. Solid black lines are the 10-order polynomial fittings. The intersections of those black lines with the y-axis give the saturation values of ${\langle t_f \rangle}_r$.}
    \label{extrapolation fig}
\end{figure}

\section{FDT computation using extrapolation\label{extrapolation}}

The mean FDT needs to be calculated using Eq.~\eqref{t mean} for protocols MPR and adaptive two-stage MPR,
which involves a summation over $n$, where $n$ runs from $1$ to $\infty$. Unfortunately, we do not have a simple expression for mean FDT like the one we have for the protocol IPR (see Eq.~\eqref{t mean x0}), hence, the
data presented in Fig.~\ref{t-mean} for protocols MPR and adaptive two-stage MPR are extrapolated data. Here, we explain the extrapolation technique in detail. Instead of computing Eq.~\eqref{t mean}, we evaluate, ${\langle t_f\rangle}_r=\langle n\tau\rangle=\tau\sum_{n=1}^{n_c}nF_{n}(r)$. 
In general, ${\langle t_f \rangle}_r$ increases with $n_c$. Figure.~\ref{extrapolation fig} shows the behaviour of  ${\langle t_f \rangle}_r$ with  $1/n_c$. We fit the data with a $10$ order polynomial of $1/n_c$. Given we are interested in the $n_c\to \infty$ result, one can read the intersection of the fitted polynomial with the $y$ axes as the extrapolated value of ${\langle t_f \rangle}_r$. The black solid lines in Fig.~\ref{extrapolation fig} represent the polynomial fittings. Figure.~\ref{extrapolation fig} (main panel) shows the data for $r=12$ (corresponds to $\Delta=5$, which gives rise to a local minima in ${\langle t_f \rangle}_r$ vs $r$ plot in Fig.~\ref{t-mean}). Inset of Fig.~\ref{extrapolation fig} shows the result for $r=16$. As one can naively expect, the convergence of the data for $r=12$ is much faster compared to  $r=16$ result. We also like to emphasize that we have checked the robustness of the extrapolated data by fitting different orders of the polynomials as well. 

\begin{figure}
    \centering
    \includegraphics[width=0.465\textwidth, height=0.3\textwidth]{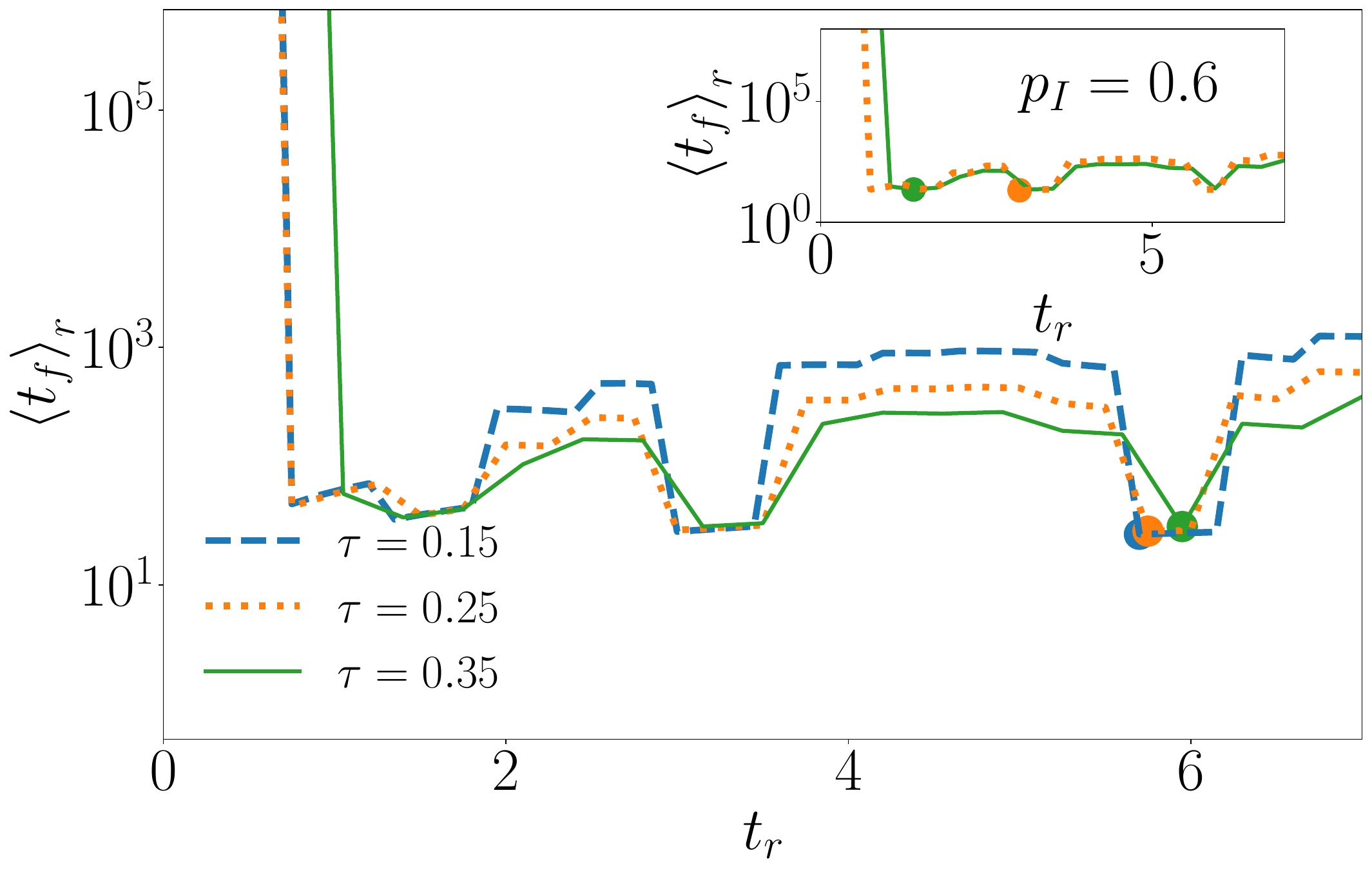}
    \caption{${\langle t_f \rangle}_r$ vs. restart time $t_r$ plots for different $\tau $ values for protocol ${\mathcal{S}}^p_{x_M}$ in main panel and for protocol $\mathcal{S}^{p_I,p_F}_{x_M}$ in inset. Circles represent the values of optimal restart time $t_r^*$.}
    \label{ tau depend}
\end{figure}

\section{$\tau$ dependence on Mean FDT\label{tau dependence}}
In this section, we check the dependency of optimal restart time $t_r^*$ on $\tau$ for protocols MPR and adaptive two-stage MPR, while we know for protocol IPR it is independent of $\tau$ in the small $\tau$ limit\cite{yin2023restart}. Figure.~\ref{ tau depend} main panel shows the variation of ${\langle t_f\rangle}_r$ with restart time $t_r$ for three different $\tau$ values, $\tau=0.15, 0.25$, and $0.35$ for protocol MPR. One can see the optimal values of the restart time $t^*_r$,  for all the $\tau$ values are more or less the same as represented by the circles. On the other hand, inset of Fig.~\ref{ tau depend} shows that for the protocol adaptive two-stage MPR, the optimal restart time $t_r^*$ is sensitive to the value of $\tau$, as on changing $\tau$ the optimal restart time switches from one local minima to the another local minima. 

\textcolor{black}
{\section{Evolution under repetitive projective measurements}\label{Appendix-C}}
\begin{figure}
    \centering
    \includegraphics[width=0.465\textwidth, height=0.3\textwidth]{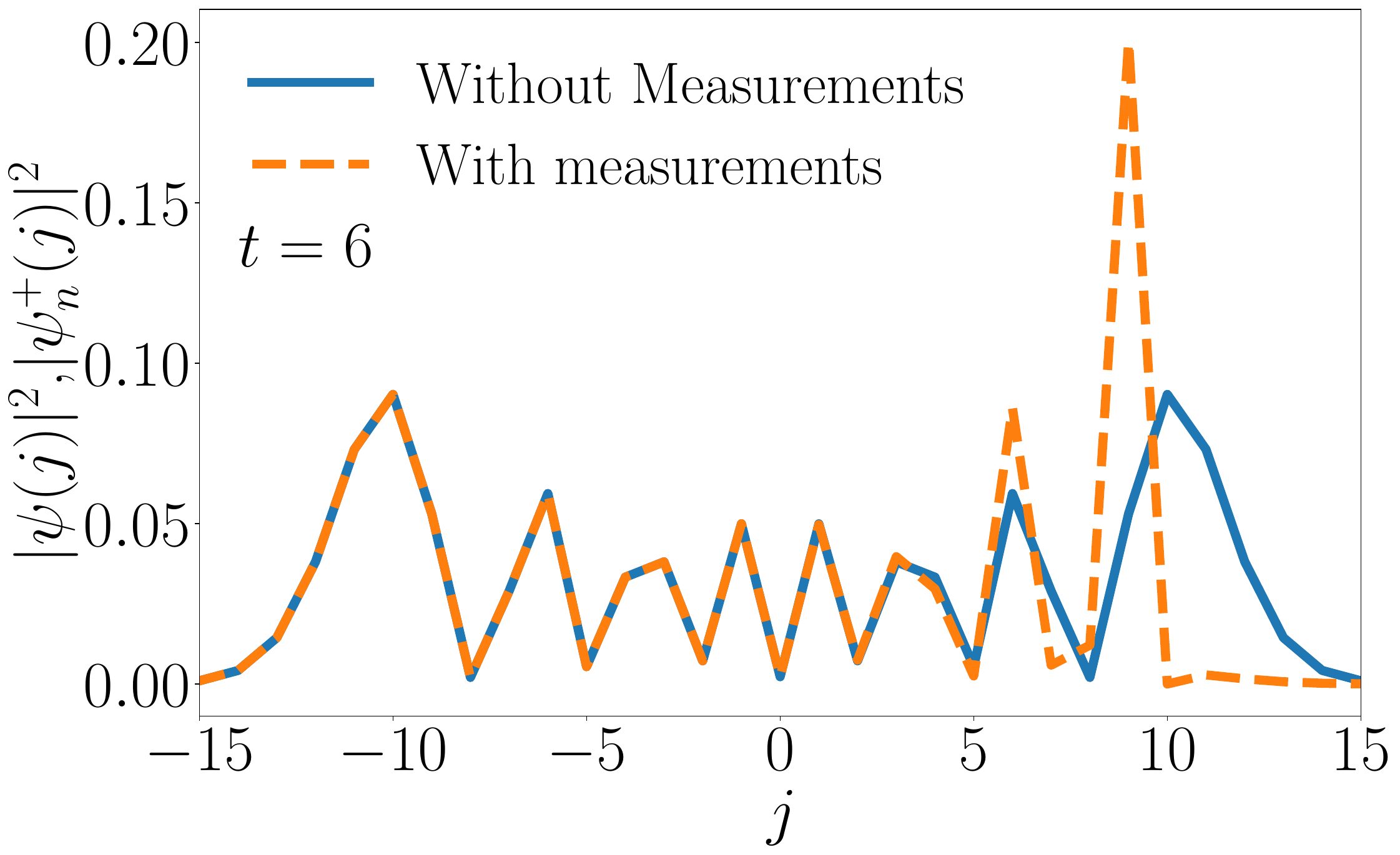}
    \caption{$|\psi(j)|^2$ vs. $j $ for measurement-free unitary dynamics (solid line) and $|\psi_n^+(j)|^2$ vs. $j $ plot for a dynamic that gets interrupted by $n=24$ numbers of projective measurements (dashed line). Results are for $t=6$ ($\Delta=10$), $\delta=10$, and $\tau=0.25$, and the initial position of the particle is at the site $j=0$.}
    \label{MP_r_24 }
\end{figure}

\textcolor{black}{If the dynamics of a particle staring from a site $x_0$, is interrupted by  the repetitive projective measurements at the site $\delta$,  the wave function { (unnormalized)}  immediately after the $n$-th measurement(given measurements does not detect the particle) is given by,
\begin{equation}
    |\psi_n^+\rangle=[(I-|\delta\rangle\langle\delta|)\hat{U}(\tau)]^n|x_0\rangle.
\end{equation}
}
\textcolor{black}{Figure.~\ref{MP_r_24 } shows the time evolution of the norm square of the wave function starting from the site $j=0$  interrupted by projective measurements at site $\delta=10$.}  
\textcolor{black}{\section{Numerical experiment and validation 
 of Eq.~\eqref{new F_n}}\label{Appendix-D}
\begin{figure}
    \centering
    \includegraphics[width=0.465\textwidth, height=0.3\textwidth]{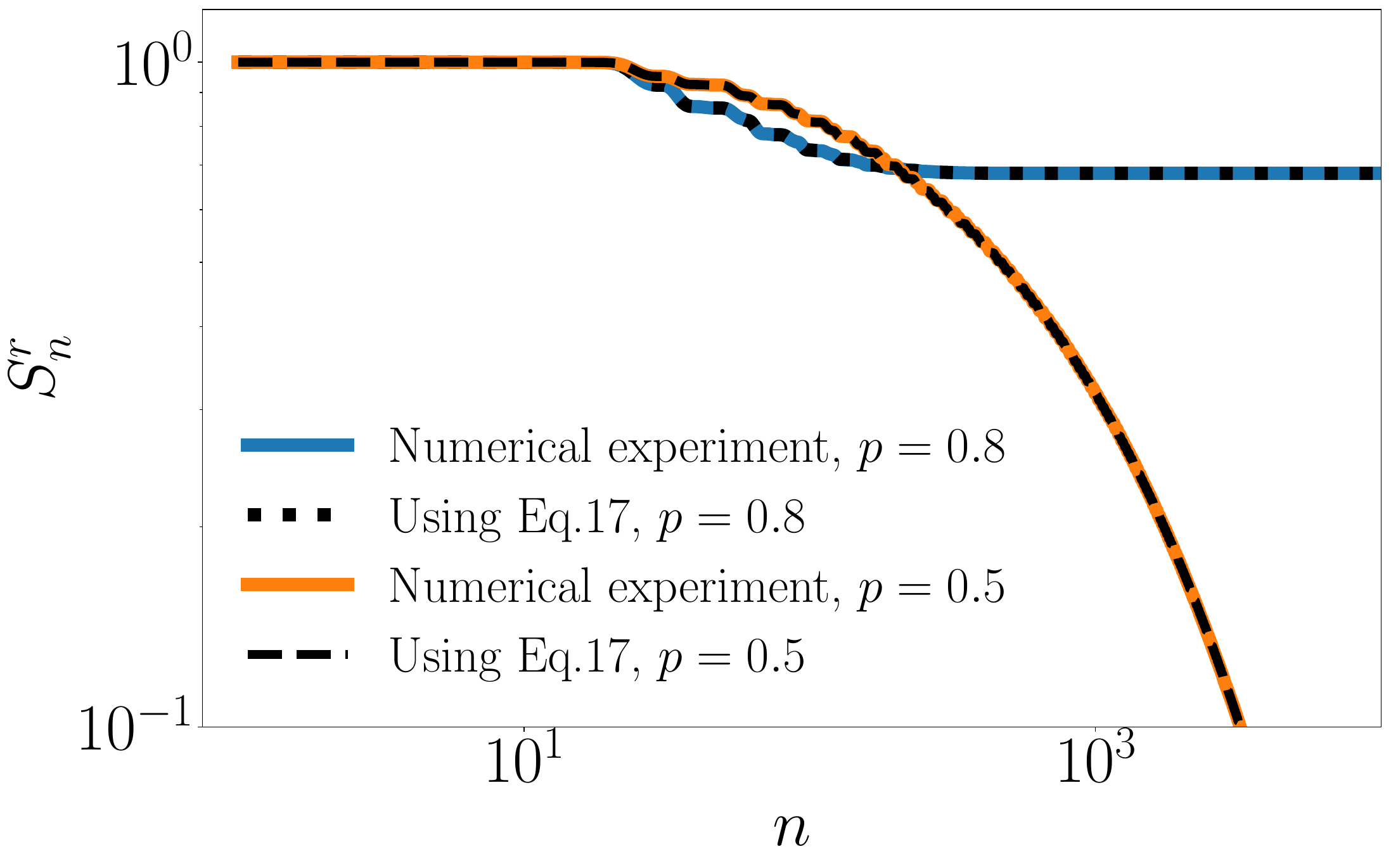}
    \caption{$S^r_n$ vs. $n$ plots. Solid lines represent the values of survival probabilities calculated numerically using weighted random numbers for different $p$ values. The dashed line represents the value of survival probability calculated from Eq.~\eqref{new F_n} for $p=0.5$ and the dotted line represents the value of survival probability calculated from Eq.~\eqref{new F_n} for $p=0.8$. Results are for $r=15$.}
    \label{S_comp }
\end{figure}
In this section, we present the results of a numerical experiment to prove the correctness of  Eq.~\eqref{new F_n}. 
Assuming $p$ ($q$) is the probability of resetting to the right (left) side of the lattice and if we start from an initial position $x_0$, then in the 1st step, the probability of resetting at $x_0+\Delta$ and $x_0-\Delta$ will be $p$ and $q$ respectively  (assume $\Delta >0$). So if we follow this protocol, say for $N$ times, for large $N$, roughly, we will reset the particle $Np$ and $Nq$ times at $x_0+\Delta$ and $x_0-\Delta$, respectively. 
In the 2nd step,  the probability of resetting at $x_0+2\Delta$, $x_0$, and $x_0+2\Delta$ will be $p^2$, $2pq$, and $q^2$ respectively independent of where the particle has been reset in the previous step. It means,  roughly, we will reset the particle $Np^2$, $2Npq$, and $Nq^2$ times at $x_0+2\Delta$, $x_0$, and $x_0-2\Delta$, respectively independent of the particles resetting position in the previous step. We keep on resetting the particles in subsequent steps in appropriate positions with appropriate probabilities. Following this process, we can generate the mean first detection probability for each realization and take an average of $N=5000$ realizations to get the final mean first detection probability of $F_n(r)$. Using this final mean first detection probability, we calculate the survival probability $S^r_n$ for $r=15$ and compare this result with the result obtained directly from Eq.~\eqref{new F_n} and Eq.~\eqref{binom}. Figure.~\ref{S_comp }  clearly shows the results obtained from the numerical experiments and from Eq.~\eqref{new F_n} are in very good agreement for $p=0.5$ and $p=0.8$ cases. The solid lines represent the results of $S_n^r$  from the numerical experiment, and the dotted and dashed lines correspond to the results obtained using Eq.~\eqref{new F_n} and Eq.~\eqref{binom}.
It indeed validates the correctness of Eq.~\eqref{new F_n} for path-independent MPR protocols. 
}

\textcolor{black}{\section{Typical mean FDT for path-dependent MPR protocol }\label{Appendix-E}
\begin{figure}
    \centering
    \includegraphics[width=0.465\textwidth, height=0.3\textwidth]{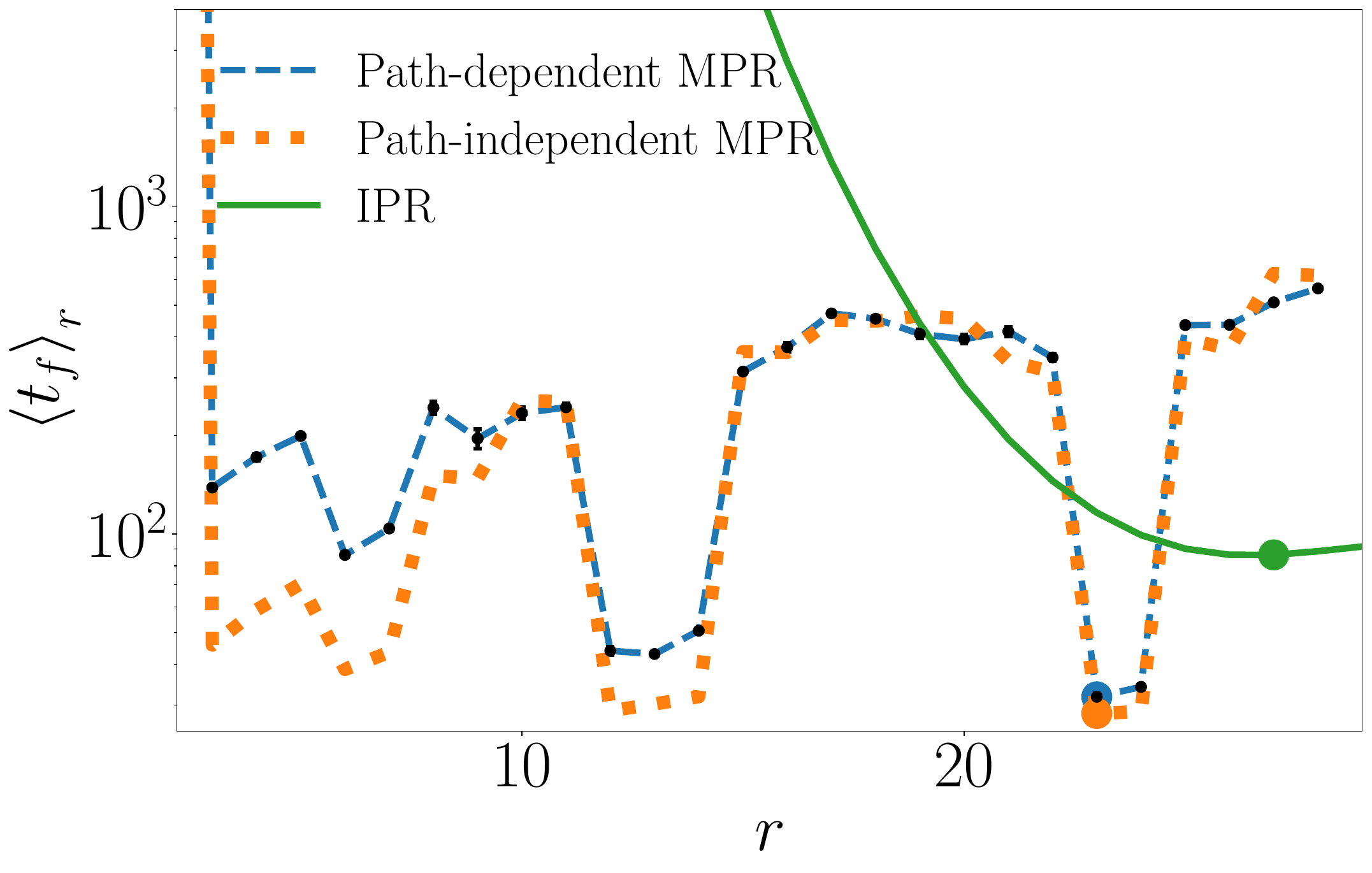}
    \caption{ Mean ${\langle t_f\rangle}_r$ vs. resetting rate $r$ plot. The blue dashed and orange dotted lines represent the typical mean FDT for the path-dependent and mean FDT for the path-independent (one described in the main text) MPR protocol. The solid line represents the IPR protocol.}
    \label{t_typical }
\end{figure}
In the main text, we have considered the path-independent MPR protocol. Here, we also demonstrate the results for path-dependent MPR protocol. 
In the case of path-dependent MPR protocol (where one tracks the path of the successive resetting positions), the First detection probability 
 of a given reset sequence can be written as,
\begin{equation}
    F_n^{\{\sigma_i\}_{i=0}^R}(r)=F_{\Tilde{n}}^{j_R}\prod_{i=0}^{R-1}(1-P^{j_i}_{det}(r)),
\end{equation}
where $\sigma_i=\pm 1$, $j_0=x_0$,  $j_{i>0}=j_{i-1}+\sigma_i\Delta$, $F_n^{j_i}=|\phi_n^{j_i}|^2$ with $\phi_n^{j_i}$ the
amplitude at $\delta$ under the detection process started from position $j_i$ and
followed by $n-1$ failed detections, and $P^{j_i}_{det}(r)=\sum_{n=1}^r F_n^{j_i}$. \\}

\textcolor{black}{The probability of a given reset sequence is given by, 
\begin{equation}
    P\big(\{\sigma_i\}_{i=0}^R\big)=\prod_{i=1}^R\large[p\frac{1+\sigma}{2}+(1-p)\frac{1-\sigma}{2}\big].
\end{equation}
Hence, the first detection probability after $n$th attempt of detection can be obtained from,
\begin{equation}
    F_n(r)=\sum_{\{\sigma_i\}_{i=0}^R} P\big(\{\sigma_i\}_{i=0}^R\big)F_n^{\{\sigma_i\}_{i=0}^R}(r).
\end{equation}}

\textcolor{black}{We have considered a large number of realizations (we have checked the convergence of our data by changing the number of realizations) and numerically calculated conditional mean FDT~\cite{wang2024firstnew} for each possible path. Then, we calculate a typical average of mean FDT over those large numbers of realizations.
In Fig.~\ref{t_typical }, the blue dashed line represents the typical mean FDT for different restart rates $r$ for path-dependent MPR protocol, while the orange dotted line corresponds to the path-independent MPR protocol described in the main text (note that data presented in Fig.~\ref{t_typical } are extrapolated one, as we have described in Appendix.~\ref{extrapolation}). The data suggests that the mean FDT corresponding to the optimal resetting rate for the path-independent protocol is a bit smaller than the path-dependent one, making our path-independent protocol described in the main text more efficient.   
}
\textcolor{black}{\section{Relation between $t_r$ and $\Delta$}\label{Appendix-F}}
\begin{figure}[!h]
    \centering
    \includegraphics[width=0.465\textwidth, height=0.3\textwidth]{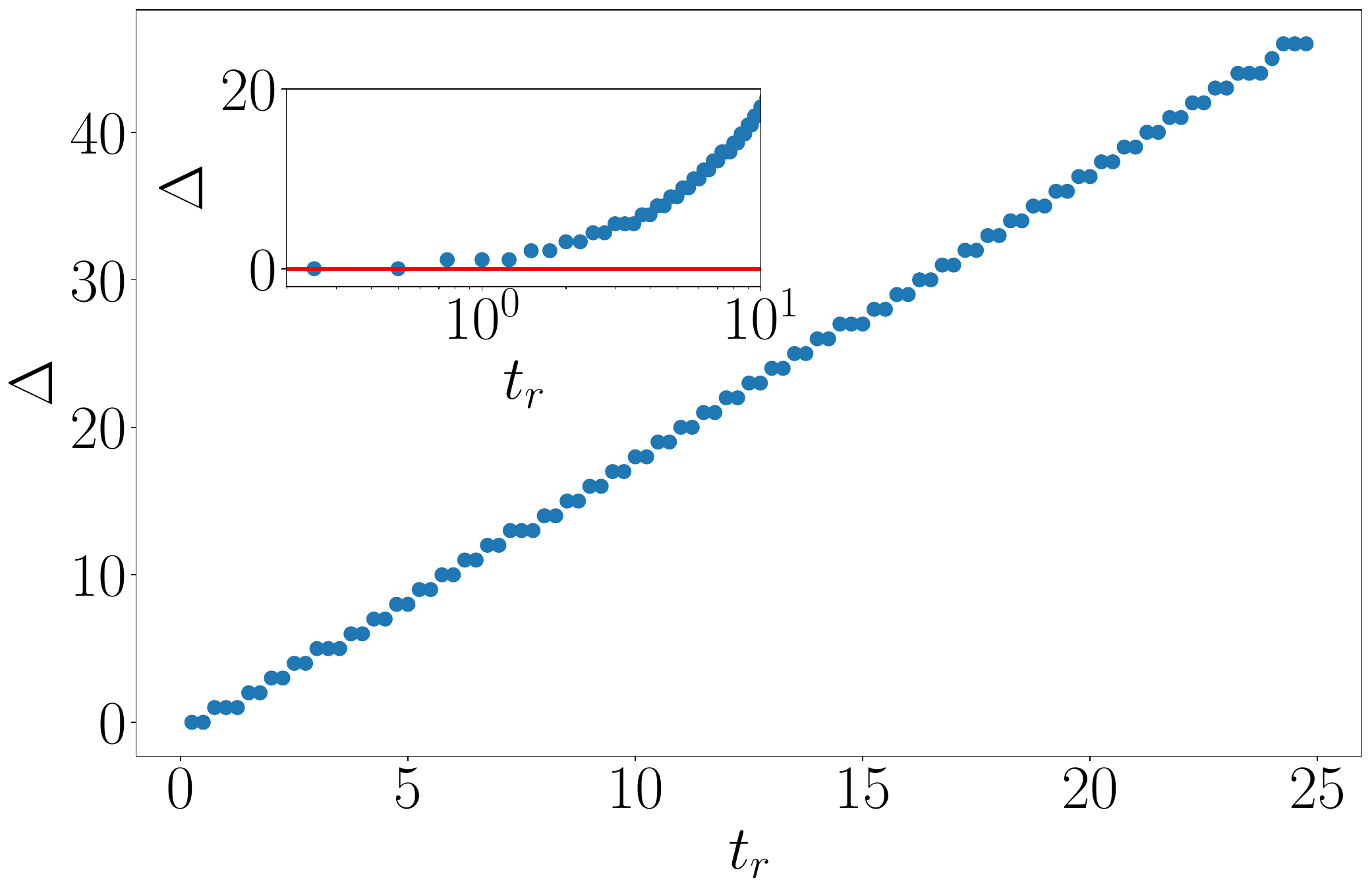}
    \caption{$\Delta$ vs. $t_r$ plot for $\tau=0.25$.}
    \label{Delta vs t_r}
\end{figure}
\textcolor{black}{Figure.~\ref{Delta vs t_r} shows the relation between $\Delta$ and $t_r$ for $\tau=0.25$. While it seems $\Delta$ increases linearly with $t_r$, but $\Delta$ remains zero for $t_r<3\tau$. It implies that below the critical resetting time $t^c_r=0.75$, IPR and MPR protocols are equivalent.}
\bibliography{reset}
\end{document}